\documentclass{article}
\usepackage[utf8]{inputenc}
\usepackage{geometry}
\usepackage{amssymb}
\usepackage{amsmath}
\usepackage{cases}
\usepackage{graphicx}
\usepackage[dvipsnames,table]{xcolor}
\usepackage{listings}
\usepackage{unicode-math}
\usepackage{titling}
\usepackage{float}
\usepackage{array}
\usepackage{makecell}
\usepackage{todonotes}
\usepackage{subcaption}
\usepackage{fontspec}
\usepackage{soul}
\usepackage[normalem]{ulem}
\usepackage{caption}
\usepackage{mathrsfs}
\usepackage{booktabs}
\usepackage{hyperref}

\usepackage[ruled,vlined,nofillcomment]{algorithm2e}

\usepackage[numbered,framed,autolinebreaks,useliterate]{matlab-prettifier}

\usepackage[
  maxbibnames=10,
  backend=biber,
  sorting=none,
  autocite=superscript,
  citestyle=numeric,
  bibstyle=authoryear,
  giveninits=true
]{biblatex}
\DeclareNameAlias{author}{family-given}

\AtEveryBibitem{[\printfield{labelnumber}]\addspace}
\usepackage{paralist}
\usepackage{setspace}
\usepackage{enumitem}

\usepackage{MnSymbol}
\usepackage{mathtools}
\usepackage{scrextend}
\usepackage[right]{lineno}
  \usepackage[overload]{empheq}
    \usepackage{cases} 

\definecolor{myblue}{RGB}{0,76,153}
\definecolor{mygreen}{RGB}{0,120,0}
\definecolor{mygray}{gray}{0.95}

\definecolor{codeframe}{gray}{0.70}
\definecolor{codecapbg}{gray}{0.20}
\definecolor{codecapfg}{gray}{0.99}

\usepackage{lstbayes}

\usepackage{caption}
\DeclareCaptionFont{codecapfg}{\color{codecapfg}}
\DeclareCaptionFormat{listing}{%
  \hspace*{-0.4pt}
  \colorbox{codecapbg}{%
    \parbox{\dimexpr\linewidth-2\fboxsep+0.8pt\relax}{#1#2#3}%
  }%
}
\lstdefinestyle{stan}{
language=Stan,
backgroundcolor=\color{white},   
basicstyle=\footnotesize\ttfamily, 
breakatwhitespace=false,         
breaklines=true,                 
captionpos=b,                    
commentstyle=\color{mygreen},    
extendedchars=true,              
frame=tb,                        
keepspaces=true,                 
framerule=0.3pt,
keywordstyle=\color{blue},       
numbersep=5pt,                   
numberstyle=\tiny\color{mygray}, 
rulecolor=\color{black},         
showspaces=false,                
showstringspaces=false,          
showtabs=false,                  
stepnumber=2,                    
stringstyle=\color{mymauve},     
tabsize=2,                       
}

\lstdefinestyle{Rconsole}{
language=R,
basicstyle=\ttfamily\footnotesize,
keywordstyle=,
commentstyle=,
stringstyle=,
showstringspaces=false,
breaklines=true,
columns=fullflexible,
keepspaces=true,
frame=single,
framerule=0.3pt,
rulecolor=\color{gray!40},
backgroundcolor=\color{gray!3},
moredelim=**[is][\color{blue}]{@}{@},
captionpos=b,                   
}

\usepackage{subcaption}
\begin{document}

\begin{titlingpage}
\begin{center}
\begin{Large}
A practical introduction to ODE modelling in Stan for biological systems
\end{Large}
\end{center}
\vspace{1cm}

\noindent Sara Hamis$^1$, John Forslund$^1$, Cici Chen Gu$^1$, Jodie A. Cochrane$^1$.
\vspace{0.5cm}

\small{
\noindent $^1$Department of Information Technology, Uppsala University, Uppsala, Sweden.
}
\vspace{0.5cm}

\noindent Correspondence: sara.hamis@it.uu.se. 
\vspace{.5cm}

\noindent \textbf{Author ORCIDs:} SH (0000-0002-1105-8078); JF (0009-0006-5457-4298); CCG (0009-0008-1998-6486); JAC (0000-0001-8259-0478).
\vspace{.2cm}


\vspace{6cm}

\begin{abstract}
\noindent Integrating dynamical systems models with time series data is a central part of contemporary mathematical biology. 
With the rich variety of available models and data, numerous methods and computational tools have been developed for these purposes. 
One such tool is Stan, a freely available and open-source probabilistic programming framework that provides efficient methods for estimating model parameters from data using computational Bayesian inference algorithms. 
Stan includes built-in mechanisms for working with ordinary differential equation (ODE) models, which are widely used in mathematical biology and related fields to study simulated, experimental, and real-world systems that change over time. 
Through step-by-step worked examples, including both pedagogical toy models and applications with real data, this article provides a practical, self-contained  introduction to performing parameter estimation and model evaluation for first-order linear and nonlinear ODE models in Stan. 
The article also explains key statistical methods that underpin Stan and discusses computational Bayesian modelling in the context of biological applications. 

\end{abstract}

\end{titlingpage}

\newpage
\vfill
\section{Introduction}
With the recent surge in biological data, integrating mathematical models with empirical observations has become a central part of contemporary mathematical biology.
This trend also aligns with the increasing role of model-informed practices in medicine, public health, and environmental sectors \cite{Pepin2025}. 
However, given the rich variety of models and data used in the field, there is no consensus on best practices for parameterising, evaluating, and selecting models with respect to data. 
Instead, a variety of methods and computational tools have been developed for these purposes. 
This article focuses on implementing a particular family of models, ordinary differential equation (ODE) models, within the context of probabilistic modelling of dynamical systems in biology, using one such tool: Stan~\cite{stan}.
Stan is a non-proprietary probabilistic programming framework that provides an efficient way to estimate model parameters from data using computational Bayesian inference algorithms.

Given the myriad of available methods for parameter estimation, model evaluation, and model selection, why should a reader then (a) use Stan and (b) read this article?
To answer question (a), let us first acknowledge that Stan makes sophisticated Bayesian parameter estimation methods available to researchers who are not specialists in computational statistics. 
This allows Stan users to focus on other aspects of their research, while Stan is continuously developed and maintained by a dedicated team of experts. 
Moreover, Stan is open-source and well-documented, making its underlying algorithms transparent and enabling custom extensions of the code base. 
Despite this, Stan remains underused in mathematical biology. 
This claim is supported by a citation-data analysis we conducted, revealing that only 15 articles published between Stan’s release in 2012 and 2024 cite Stan across five well-established mathematical biology journals \cite{pugh_bib_2025} (Supplementary Material, SM1). 
This likely reflects the broad trend that frequentist statistics is more widely taught than Bayesian statistics in undergraduate mathematics and statistics programmes \cite{Hoegh2020}, and that frequentist methods remain more common than Bayesian methods in biomedical research \cite{Ferriera2020,Muehlemann2023}. 
In line with this argument, recent results from a survey-based study addressing the question {\it ``Why are there not more Bayesian clinical trials?''} suggested that insufficient knowledge of Bayesian methods is one of the main factors hindering their broader implementation \cite{Clark2023}. 
The 323 respondents were affiliated with pharmaceutical companies (33.4\%), clinical research organizations (29.7\%), regulatory agencies (18.6\%), academia, medical practice, and other. 
The under-use of Stan in mathematical biology, in particular, may also partly stem from a lack of literature resources that concisely explain how to apply Stan to common types of mathematical biology models in a way that is tailored to researchers in the field. 
Motivated by this conjecture and the prevalence of ODE models in mathematical biology, this article provides a practical guide to implementing ODE models in Stan, which answers question (b). 

In its short and practical format, our article is not designed to give readers a full background on ODE models, Bayesian statistics, or Stan. Good resources for these topics already exist, several of which are referenced throughout this article. 
Instead, our article is intended to be a streamlined introduction to implementing ODE models in Stan. 
Through worked Stan examples, we will show how to implement ODE models for biological systems whose dynamics are described by one or more linear or nonlinear first-order equations.
We will also discuss when and how to use complete, no, and partial pooling approaches, and how to perform model evaluation and selection using posterior predictive checks and estimates of the expected log predictive density. 
These techniques are conceptually explained in the Background section, and practically demonstrated with code exampled in the Implementation section.
Although this article provides a ``shortcut to implementation'', it is not intended to replace thorough background reading on the statistical theory that underlies Stan. 
On the contrary, we hope that once readers see how powerful and easy-to-use Stan is, they will be encouraged to learn more about computational Bayesian statistics. 
A friendly starting point for this learning is Lambert's introductory book on Bayesian statistics, which includes worked Stan examples \cite{Lambert2018}. 
A more in-depth, although still introductory, discussion on the methods on which Stan is built is provided by \textcite{betancourt2017conceptual}.
In addition, specialised resources on how to use Stan in specific research areas of mathematical biology include articles on pharmacometrics \cite{Margossian2022}, epidemiology \cite{Grinsztajn2021,Chatzilena2019}, and ecology \cite{Fraenzi2015}. 
\vfill

\section{Background}

\subsection{Back to basics and Bayes' rule}
\label{sec:bayes_rule}
\begin{figure}[h!]
    \centering
    \includegraphics[width=1\linewidth]{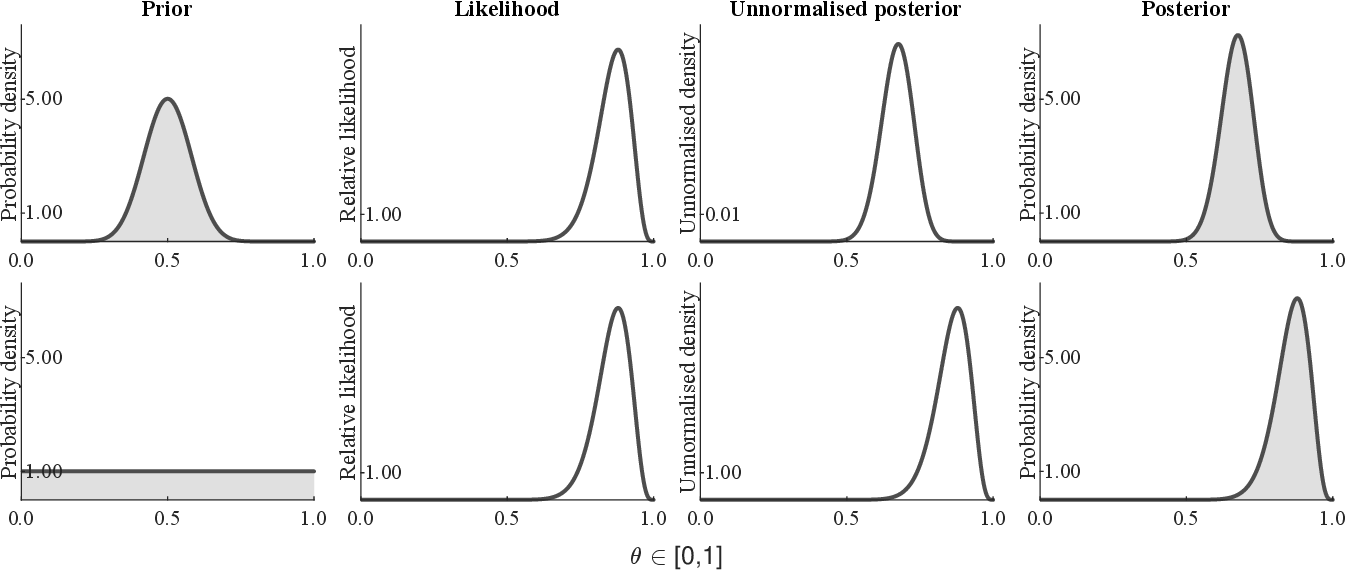}
    \caption{
    {\bf A graphical illustration of Bayes’ rule with informative and uninformative priors.}
    The top row has an informative prior $p(\theta)=\mathrm{Beta}(20,20)$, which is approximately Normal in shape and centered at $\theta=0.5$, while the bottom row has an noninformative prior $p(\theta)=\mathrm{Beta}(1,1)$, equivalent to a uniform distribution on $[0,1]$.
    In both cases, the likelihood takes the shape of a $\mathrm{Beta}(30,5)$ distribution as a function of $\theta$ and is shown on a relative scale, since it represents the probability of the observed data conditional on $\theta$, rather than a density over $\theta$ itself.
    The unnormalised posteriors are obtained by pointwise multiplication of the prior and likelihood in $\theta$.
    These are normalised in the right-most column, yielding posteriors that are proper probability distributions.
    Grey-shaded areas under the curves represent probability density functions normalised to unit area, whereas unshaded curves represent quantities shown on relative (unnormalised) scales.
    }
    \label{fig:1}
\end{figure}

Our starting point for Bayesian parameter inference is Bayes' rule in the form

\begin{equation}
     p(\theta \mid \mathcal{D}) = \frac{p(\mathcal{D} \mid \theta)p(\theta)}{p(\mathcal{D})},   
     \label{eq:BayesFormula}
\end{equation}
where $\mathcal{D}$ denotes the observed data, and $\theta$ denotes the model parameter(s) whose values we want to infer. 
If we have $j$ such parameters, then $\theta$ is a vector $\theta=(\theta_1,\theta_2,...,\theta_j$). 
Eq.~\eqref{eq:BayesFormula} relates the posterior distribution $p(\theta \mid \mathcal{D})$ to three right-hand side components: the likelihood function $p(\mathcal{D} \mid \theta )$, the prior distribution $p(\theta)$, and the normalising constant $p(\mathcal{D})$. 
Fig.~\ref{fig:1} conceptually illustrates how these components combine to form the posterior.

The main goal of Bayesian inference is to learn the posterior distribution $p(\theta \mid \mathcal{D})$, 
which represents what we believe about the parameter values {\it after} letting data inform our model.
In contrast, the prior $p(\theta)$ represents our beliefs about the parameter values {\it before} this data-informed update.
For some applications in mathematical biology, we can leverage knowledge about the modelled system and use informative priors. In other cases, such knowledge is unavailable, and noninformative or weekly informative priors could be more appropriate \cite{gelman1995bayesian}. 
The effect of using informative and noninformative priors is exemplified in Fig.~\ref{fig:1}. 

We next consider the likelihood $p(\mathcal{D} \mid \theta )$, which describes how likely it is to have observed the data $\mathcal{D}$, given a specific model and value of $\theta$.
Note that in our setting, the likelihood incorporates both the ODE model and the error model describing either measurement noise or biological stochasticity.
Lastly, we remark that the numerator in Eq.~\eqref{eq:BayesFormula} generally does not integrate (for continuous $\theta$) or sum (for discrete $\theta$) to one, and therefore does not by itself define a valid probability distribution.
Dividing by the denominator normalises the right-hand side and ensures that the posterior is a valid probability distribution. However, as we will see later (in Section~\ref{sec:bayestostan}), it is not necessary to compute $p(\mathcal{D})$ explicitly when using the computational Bayesian inference methods that Stan builds upon.
\vspace{1cm}

\subsection{Embracing uncertainty in Bayesian statistics and mathematical biology}
When we estimate model parameters with Bayesian inference methods, we obtain posterior distributions that span a range of parameter values and express our uncertainty about them. 
This probabilistic representation transparently demonstrates that Bayesians associate uncertainty with model parameters and, by extension, with models themselves. 
This uncertainty can be understood through two contrasting interpretations (A) and (B):
\begin{enumerate}[label=(\Alph*)]
    \item The biological process that generated the data can be described by a mathematical model with deterministic parameters. However, we are uncertain about the true values, and this epistemic uncertainty is reflected in the parameter distributions.
    \item The biological process that generated the data can be described by a mathematical model with 
    parameter values that vary across biological units according to an underlying distribution. 
    In this interpretation, 
    Bayesian inference characterises both biological variability and uncertainty about the corresponding parameter distributions. 
\end{enumerate}
\noindent Both interpretations (A) and (B) have merit in mathematical biology. Importantly, although they differ philosophically, both can practically be handled using the same statistical methodology. We can therefore proceed with the technical aspects of Bayesian inference without a philosophical digression.
That said, for a broader discussion of how uncertainty relates to scientific practice, see Kampourakis and McCain's book \cite{KampourakisMcCain2019}.

\subsection{What (and who) is Stan?}
Stan is a probabilistic programming language designed for Bayesian inference \cite{stan}. 
The name itself pays tribute to Stanis\l{}aw Ulam, one of the mathematicians behind the Monte Carlo method \cite{MetropolisUlam1949}. 
In line with this, Stan builds on advanced Monte Carlo algorithms to efficiently compute posterior distributions given user-specified models and data. 
To interface with Stan, we have to work in a computing environment from which we can pass data and handle post-sampling analysis. 
Supported environments include interfaces for popular programming languages such as R, Python, Julia, MATLAB, and Mathematica. 
In the implementation parts of this article, we will work with RStan, which gives access to multiple R packages that streamline model analysis and evaluation. 
Due to its popularity, online forums provide ample support for RStan.
In addition, the integrated development environment RStudio includes several helpful features for writing Stan code, such as colour-coded typesetting, auto-completion, and inline diagnostics, making it an excellent environment for getting started with Stan.

\subsection{From Bayes' rule to Stan's algorithms}
\label{sec:bayestostan}
Bayes’ rule is strikingly elegant, but there are several steps between understanding it in its basic form (Eq.~\ref{eq:BayesFormula}) and implementing it in computer code for ODE models. 
We will walk through these steps in Sections~\ref{sec:bayestostan} and \ref{sec:bayesODE}, building up both intuition and the statistical foundations needed to understand Stan-based inference.

\subsubsection{Sampling from posterior distributions}
While the numerator of Eq.~\eqref{eq:BayesFormula} is straightforward to evaluate, the denominator often becomes analytically intractable when $\theta$ is high-dimensional and continuous, as it requires integrating the likelihood–prior product over the entire parameter space, which typically has no closed-form solution. 
For certain likelihood–prior pairs, e.g., as conjugate priors, the posterior can be obtained in closed form \cite{murphy2007conjugate}. 
However, in most biological applications, restricting the analysis to conjugate priors is both impractical and biologically unmotivated, so approximate methods are required to characterise the posterior distribution. 
A range of posterior approximation techniques have been developed, including variational inference \cite{jordan1999introduction}, Laplace approximation \cite{Tierney1986}, and related deterministic approaches such as expectation propagation \cite{Minka2001}. 
In this work, we focus on sampling-based inference methods built on Markov chain Monte Carlo (MCMC) \cite{Tierney1994}, which underpin Stan's algorithms.

As a first practical step toward sampling, we observe that since the relative values of the posterior density determine how likely different parameter values are, we do not need to normalise the posterior to reason about which values are more or less probable. 
Therefore, we can work with Bayes' rule in its simplified, proportional form
\begin{equation}
     p(\theta \mid \mathcal{D}) \propto p(\mathcal{D} \mid \theta )p(\theta)
     \label{eq:BayesFormula_prop}
\end{equation}
\noindent and thereby avoid the need to compute the denominator altogether.
MCMC methods exploit this proportionality by evaluating only the right-hand-side product in Eq.~\eqref{eq:BayesFormula_prop} to construct a Markov chain (i.e., a stochastic sequence of parameter values such that the next state depends only on the current one) whose stationary distribution is the normalised posterior in Eq.~\eqref{eq:BayesFormula}.
When Bayes’ rule is used for MCMC-based computation, the unnormalised density is commonly referred to as the target density $\pi$, and is given by
\begin{equation}
    \pi(\theta) = p(\mathcal D\mid\theta)p(\theta).
    \label{eq:target}
\end{equation}

\begin{figure}[h!]
    \centering
    \includegraphics[width=1\linewidth]{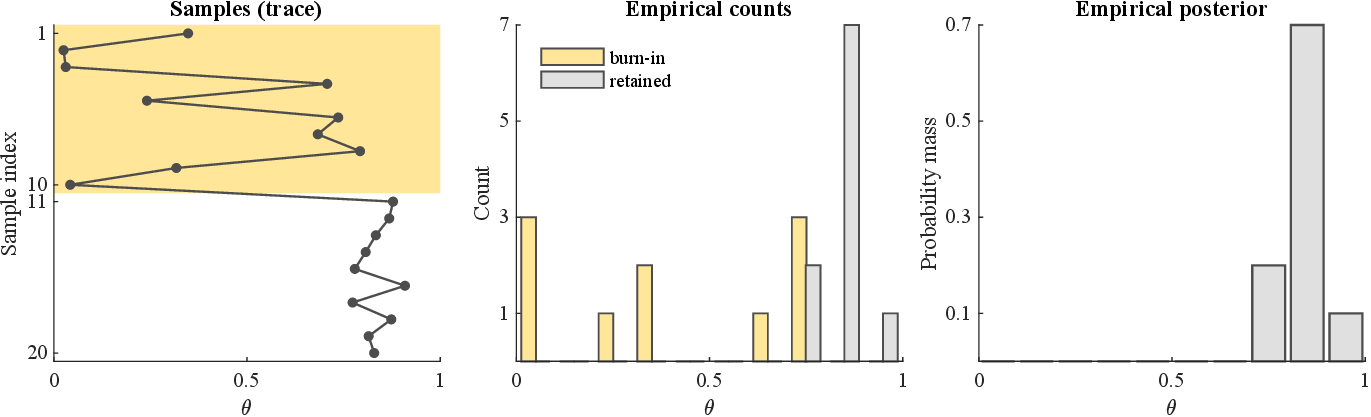}
    \caption{
    {\bf A graphical illustration of how sampling constructs an empirical posterior.}
    The left panel shows a trace of sampled $\theta$-values, 
    beginning with burn-in samples that are discarded when constructing the posterior (yellow background), followed by retained samples.
    The middle panel summarises sample counts with a binned histogram, and the right panel shows the normalised empirical posterior obtained from the retained samples, represented as a probability mass function over $\theta$. 
    }
    \label{fig:2}
\end{figure}

\noindent As a second step, we note that for continuous $\theta$ it is impossible to evaluate the posterior at all possible parameter values, and therefore aim to perform a sufficient, rather than exhaustive, exploration of the $j$-dimensional parameter space $\Theta \subseteq \mathbb{R}^j$.
The algorithm's name itself hints at how this works: ``Markov chain'' refers to the probabilistic, sequential exploration of parameter space, and ``Monte Carlo'' to the use of random sampling to approximate the posterior distribution. 
More specifically, to obtain posterior samples, MCMC algorithms construct a trajectory through $\Theta$ such that regions of high posterior density are visited more frequently than regions of low posterior density \cite{RobertCasella2004}. 
Recording the visited locations and constructing an empirical distribution from them (akin to a histogram) then provides an approximation of the posterior, as illustrated conceptually in Fig.~\ref{fig:2}. 
For visual clarity, the figure shows only a small number of samples; in practice, many more samples are generated. 
The samples are further divided into an initial burn-in period, during which the chain converges toward the stationary distribution and the samples are discarded, and a subsequent period in which recorded samples are retained for inference. 
Through this mechanism, 
MCMC algorithms thus leverage the fact that, even when the posterior in Bayes’ rule cannot be computed exactly, it can still be characterised by samples drawn from its proportional form in Eq.~\eqref{eq:BayesFormula_prop}. 

\subsubsection{It all comes down to proposing and accepting (or rejecting)}
The above discussion naturally leads to the question: How can we construct a Markov chain-based sampler that efficiently explores the parameter space $\Theta$ in order to characterise the posterior distribution of interest?
To approach this question intuitively, let us consider a general MCMC sampler.
Suppose that the sampler is currently at a location $\theta \in \Theta$, and then proposes a move to a new point $\theta' \in \Theta$, based on some proposal distribution $q(\cdot)$.
If the proposed move is \textit{accepted}, we transition from $\theta$ to $\theta'$ and store $\theta'$; if it is \textit{rejected}, we remain at $\theta$ and store the value $\theta$ (possibly again). 
In this way, the problem of sampling from the posterior becomes one of designing effective move proposals and accept/reject rules that generate empirical distributions representative of the true posterior.  
Thus, the key questions become: (1) how should moves in parameter space be proposed, and (2) how should these proposals be accepted or rejected? 
In the next subsections, we briefly introduce four  MCMC-based algorithms, presented in historical order and increasing sophistication, along with their proposal schemes and accept/reject rules. 
These are: 
random-walk Metropolis (RWM) \cite{metropolis1953equation},
Metropolis-–Hastings (MH) \cite{hastings1970monte}, 
Hamiltonian Monte Carlo (HMC) \cite{duane1987hybrid,neal2011mcmc},
and the No-U-Turn Sampler (NUTS) \cite{hoffman2014no} on which Stan is based. 
These algorithms share the common accept–reject structure formalised in Algorithm~\ref{alg:mcmc}, but differ primarily in how proposals are constructed and how acceptance probabilities are computed.

\newpage

 \begin{algorithm}[ht]
    \DontPrintSemicolon
    \SetAlgoLined
    \KwIn{
    Sampler location $\theta^{(i)}$ at iteration $i$ }
        1. Propose a new location 
        $\theta'^{(i)}$ based on the current location: $\theta'^{(i)} \sim q(\cdot \mid \theta^{(i)})$ \;
        2. Calculate the acceptance probability $\alpha$ of moving to the proposed location  
        \; 
        3. Accept move to $\theta^{(i+1)}=\theta'^{(i)}$ with probability $\alpha$; otherwise remain at $\theta^{(i+1)}=\theta^{(i)}$\; 
    \KwOut{
    Updated sampler location $\theta^{(i+1)}$ }
     \caption{General strategy for MCMC-based sampling methods}\label{alg:mcmc}
\end{algorithm}

\subsubsection{Random-walk Metropolis}
The RWM algorithm proposes a new position $\theta'$ by sampling from a multivariate normal distribution centred at the current position $\theta$ with covariance matrix $\Sigma$, such that
\begin{equation}
\theta' \sim \mathcal{N}(\theta, \Sigma).
\end{equation}
Here $\Sigma$ controls the proposal step sizes. 
In classical RWM, $\Sigma$ is specified \emph{a priori}; 
however, in practice it is often tuned to achieve acceptance rates close to those known to be optimal for random-walk sampling under commonly used assumptions (approximately $0.44$ for one-dimensional parameters \cite{Roberts2001} and $0.23$ in higher dimensions \cite{Roberts1997}). 
This tuning is usually performed during an initial sampling phase, in which proposal parameters are adjusted and samples are discarded.

\begin{figure}[h!]
    \centering
    \includegraphics[width=1\linewidth]{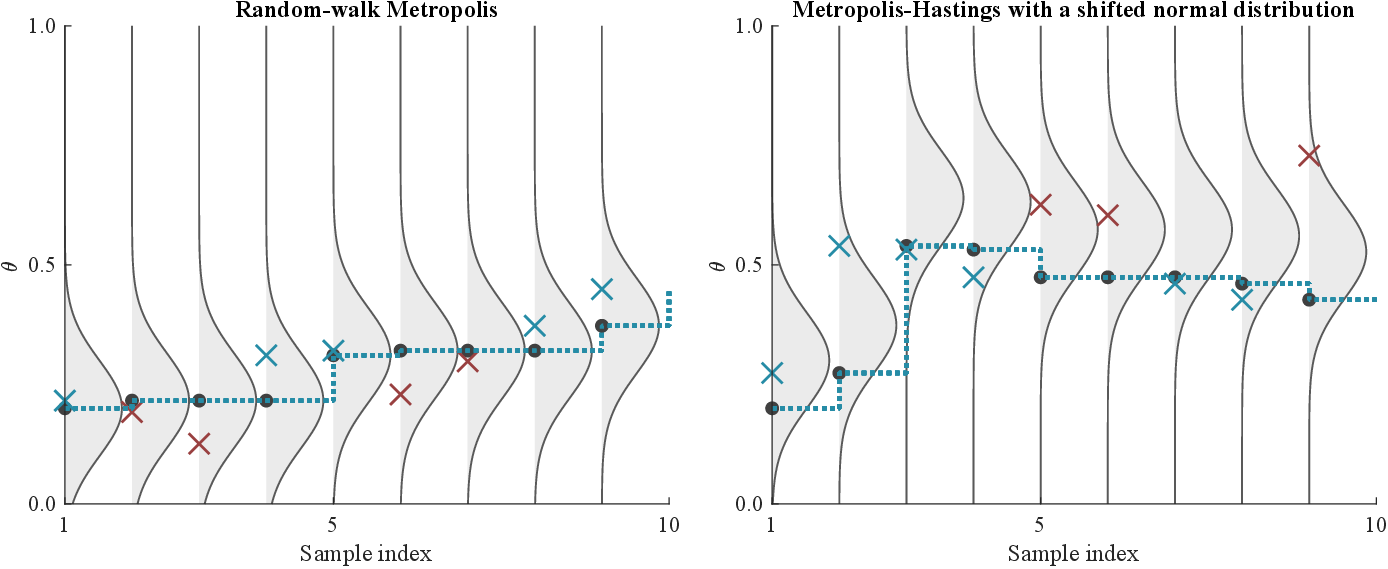}
    \caption{
    {\bf A schematic demonstration of proposal distributions in random-walk Metropolis (left) and Metropolis--Hastings with a shifted normal proposal (right).} 
    At each sample index $i$, the filled circle marks the current state $\theta^{(i)}$ and the ridge depicts the proposal density.
    The cross at index $i$ indicates the proposed state $\theta^{'(i)}$.
    At the next index $i+1$, the state satisfies 
    $\theta^{(i+1)}=\theta^{'(i)}$ upon acceptance and 
    $\theta^{(i+1)}=\theta^{(i)}$ upon rejection, as denoted by teal and red crosses, respectively. 
    The dotted teal trace connects the successive states $\theta^{(i)}$, illustrating the trajectory of the Markov chain over the sample indices.
    }
    \label{fig:3}
\end{figure}

Once the covariance matrix $\Sigma$ is fixed, 
the algorithm proceeds to accept or reject proposed samples according to the Metropolis rule. 
Specifically, a proposed move from the current location $\theta$ to a new location $\theta'$ is accepted with probability
\begin{equation}
\alpha = \min \left(1, \: \frac{\pi(\theta')}{\pi(\theta)} \right),
\label{eq:alpha_rwm}
\end{equation}
so that proposals to regions of higher posterior density are always accepted, while proposals to lower-density regions are accepted with a probability given by the posterior ratio.
This makes it explicit that the target density defined in Eq.~\eqref{eq:target} need not be normalised, since MCMC acceptance probabilities depend only on ratios of the target density evaluated at two points in parameter space, canceling out the common denominator term.
The RWM algorithm is illustrated in Fig.~\ref{fig:3}, which highlights that the proposal distribution is symmetric and independent of the local geometry of the posterior being explored. 
This makes RWM easy to understand and implement, but also means that it is often inefficient at exploring posterior spaces, especially in high dimensions.

\subsubsection{Metropolis--Hastings}
MH extends the RWM algorithm by allowing for more flexible proposal distributions, whose asymmetry is accounted for in the acceptance probability. 
This partially remedies the inefficient parameter exploration that characterises RWM. 
Specifically, in MH, the proposed step to $\theta'$ depends on the current location $\theta$ according to
\begin{equation}
\theta' \sim q(\cdot \mid \theta),
\label{eq:q_prop}
\end{equation}
where $q(\cdot \mid \theta)$ denotes the full distribution from which a proposal is sampled. 
In MH, this distribution can take any shape, be it non-Gaussian, asymmetric, or problem-specific, as exemplified in Fig.~\ref{fig:3}.
The proposed $\theta'$ is then accepted with probability 
\begin{equation}
\alpha = \min \left(1, \: \frac{\pi(\theta')\, q(\theta \mid \theta') }{\pi(\theta)\, q(\theta' \mid\theta)} \right).
\label{eq:alpha_MH}
\end{equation}

\noindent Here the ``flipped'' $q$-ratio compensates for directional asymmetries that occur if the proposal distribution is more likely to step from $\theta$ to $\theta'$ than back, or vice versa.  
Note that if $q$ is chosen to be a normal distribution centred at $\theta$ with covariance $\Sigma$, then MH simply reduces to RWM. 
In practice, the proposal distribution is chosen and tuned to balance exploration and acceptance, and can range from simple random-walk proposals to more problem-specific ones. 
While MH is more flexible than RWM, its proposals are still largely heuristic and do not exploit information about the posterior geometry, which limits efficient exploration of $\Theta$.

\subsubsection{Hamiltonian Monte Carlo}
\begin{figure}[h!]
    \centering
    \includegraphics[width=1\linewidth]{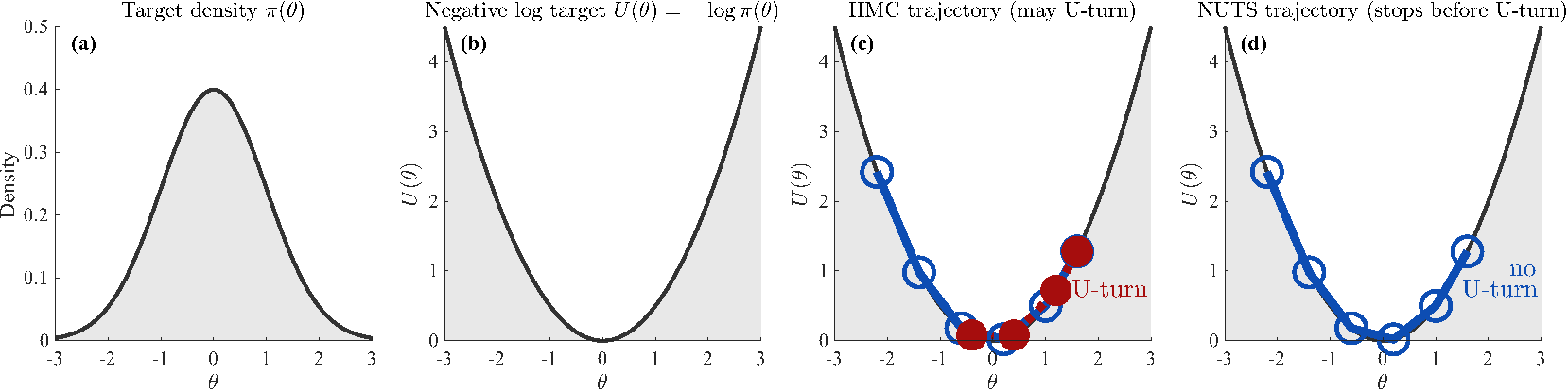}
    \caption{
    {\bf A schematic illustration of Hamiltonian Monte Carlo (HMC) and the No-U-Turn Sampler (NUTS) in a one-dimensional parameter space.}
    (a) The target distribution \(\pi(\theta)\), i.e.\ the distribution explored by the sampler.
    (b) The negative log of the target defines a potential energy landscape on which the HMC and NUTS samplers operate.
    (c) A schematic HMC trajectory on the potential energy landscape, illustrating that fixed-length trajectories may reverse direction, i.e., U-turn.
    The Hamiltonian trajectory (solid line) and leapfrog stopping points (open circles) prior to the U-turn are shown in blue and correspond to motion toward increasing \(\theta\).
    The trajectory (dashed line) and leapfrog stopping points (filled circles) after the U-turn are shown in red and correspond to motion toward decreasing \(\theta\).
    (d) A schematic NUTS trajectory, which adaptively terminates the trajectory to avoid U-turns.
    Trajectories are schematic representations of a single leapfrog-based Hamiltonian trajectory.
    }
    \label{fig:4}
\end{figure}

HMC addresses the inefficiency of MH by borrowing ideas from classical mechanics. 
In HMC we work with the target density in its negative log form, where the negation flips the target density landscape so that valleys correspond to regions of high posterior density, and the logarithm converts the likelihood--prior product in Eq.~\eqref{eq:target} into a sum, making it more convenient to evaluate numerically. 
The transformed posterior landscape can be interpreted as a potential energy landscape, as visualised in Fig.~\ref{fig:4}.
As part of its proposal mechanism and in line with the energy analogy, the HMC algorithm defines a Hamiltonian ${\cal H}$ consisting of a potential energy $U(\theta)$ and a kinetic energy $K(r)$ such that
\begin{equation}
{\cal H}(\theta,r)=U(\theta)+K(r). 
\label{eq:hamiltonian}
\end{equation}
Under this formulation, HMC augments the parameter vector $\theta$ with an auxiliary momentum variable $r$ of the same dimension, which has no statistical meaning but controls how the sampler moves through parameter space. 
The potential energy $U(\theta)$ corresponds to the height of the negative log-target density landscape, while the kinetic energy $K(r)$ determines how strongly the sampler is pushed across this landscape.
Using the negative-log construction introduced above, the Hamiltonian can be written in terms 
of the joint target density over $(\theta,r)$,
\begin{equation}
   \underbrace{-\log(\pi(\theta,r))}_{\mathcal{H}(\theta,r)}
=
\underbrace{-\log(\pi(\theta))}_{U(\theta)}
+
\underbrace{-\log(\pi(r \mid \theta))}_{K(r)},
\end{equation}
where the equality holds up to an additive constant that does not depend on $\theta$ or $r$ and therefore does not affect the dynamics.
With this joint $(\theta,r)$ formulation, marginalising over $r$ recovers the desired posterior over $\theta$, while the auxiliary momentum variables enable efficient exploration of parameter space.
\\

\noindent 
Specifically, in standard HMC, this exploration is achieved by setting
\begin{equation}
K(r) = \tfrac{1}{2} r^\top M^{-1} r,
\label{eq:K_kin}
\end{equation}
reflecting the quadratic form of kinetic energy, where $M$ is a user-defined mass matrix that controls how the sampler moves along each parameter direction and, for each proposal, the momentum is resampled according to 
\begin{equation}
r \sim \mathcal{N}(0, M).
\end{equation}

\newpage

\noindent The core idea of HMC is to construct proposals that approximately conserve the Hamiltonian energy in Eq.~\eqref{eq:hamiltonian}, mirroring true Hamiltonian dynamics in which ${\cal H}$ is exactly conserved. 
Intuitively, this energy conservation is designed to lead to efficient exploration of the negative log-posterior landscape: 
if the sampler is in a valley where the potential energy $U(\theta)$ is low, 
then it has high kinetic energy $K(r)$, allowing it to move in large strides across the valley.
Conversely, if the sampler is on or near a peak, it has low kinetic energy and moves more cautiously.
These energy-conserving proposals avoid both the inefficient behavior of unnecessarily small steps and the low acceptance rates associated with overly large steps. 

It should be noted that the improved efficiency of HMC comes at the cost of computing gradients of the log-target density, which act as forces that steer the sampler along Hamiltonian flow, i.e., trajectories for which the Hamiltonian is conserved. 
Because these trajectories cannot be solved analytically for general posteriors (intuitively, because the Hamiltonian flow cannot be explicitly mapped out), HMC approximates the Hamiltonian dynamics using a symplectic integrator, typically the leapfrog method \cite{neal2011mcmc}, with a fixed number of steps $L$. 
Such symplectic integrators update the pair $(\theta,r)$ by alternating between updating the momentum $r$ based on the gradient of the negative log-target density, and updating the position $\theta$ based on the current momentum $r$.
Formally, HMC proposes a new pair through
\begin{equation}
(\theta',r')=\Phi_{L,\varepsilon}(\theta,r),
\label{eq:leapfrog}
\end{equation}
where $\Phi_{L,\varepsilon}$ denotes $L$ leapfrog steps with size $\varepsilon$, updating both the sampler's location and the momentum. 
In practice, $L$ is chosen in advance, while $\varepsilon$ is tuned during a warmup phase and then kept constant. 

HMC acceptance rules formally follow the same asymmetry-adjusting principle as MH (Eq.~\ref{eq:alpha_MH}), 
but here their purpose is to compensate for small violations of energy conservation introduced by numerical integration errors in Hamiltonian dynamics, 
rather than to correct for asymmetric proposal distributions.
To formulate the HMC acceptance probability, we define a joint distribution over $(\theta,r)$ as
\begin{equation}
    \pi(\theta, r) \propto \exp \big( - \mathcal{H}(\theta, r) \big),
    \label{eq:HMC_pr_prob}
\end{equation}
where the negative sign ensures that low-energy states $(\theta,r)$ are more probable than high-energy states. 
Note an important subtlety here: while the Hamiltonian is approximately conserved along Hamiltonian-flow trajectories during the proposal stage, lower-energy states correspond to higher joint target probability at the acceptance stage. 
Substituting Eq.~\eqref{eq:HMC_pr_prob} into the MH acceptance rule (Eq.~\ref{eq:alpha_MH}), the proposed pair $(\theta',r')$ is accepted with probability 
\begin{equation}
\alpha = \min \left(1, \exp \big[{\cal H}(\theta,r)-{\cal H}(\theta',r')\big] \right),
\label{eq:alpha_HMC}
\end{equation}
showing that the acceptance probability depends on differences in Hamiltonian energy at the current and proposed state, such that proposals that conserve the Hamiltonian more closely are more likely to be accepted.

\newpage
In summary, HMC suppresses the random-walk behavior inherent to RWM and MH by proposing distant, high-probability moves, 
substantially improving exploration efficiency. 
However, this comes at the cost of increased computational complexity: 
HMC requires gradient evaluations of the log-target density and careful tuning of algorithmic parameters, namely the step size $\varepsilon$, the number of leapfrog steps $L$, and the mass matrix $M$, which can be challenging in practice.

\subsubsection{The No-U-Turn Sampler}
NUTS extends HMC by addressing one of its main practical limitations: the need to choose the 
number of leapfrog steps $L$ in Eq.~\eqref{eq:leapfrog}, 
which determines the trajectory length $L \times\varepsilon$. 
We find ourselves in a Goldilocks situation. If the trajectory is too short, exploration is inefficient, but if it is too long, the sampler will eventually turn back on itself (i.e., U-turn) and waste computation.
Borrowing an analogy from Lambert \cite{Lambert2018}, we can intuit that pushing the sampler too far along the posterior landscape may cause it to lose forward momentum and reverse direction as it climbs uphill, much like a sled that slows, stalls, and turns back when pushed too far up a slope (Fig.~\ref{fig:4}). 
NUTS avoids these U-turns by detecting when a reversal is about to occur and stopping the trajectory before the sampler actually turns back.
Thus, in NUTS, the number of leapfrog steps varies between iterations and ranges from 1 to a prescribed maximum $L_{\max}$.
Stan uses a dynamic HMC algorithm based on NUTS that automatically tunes both the step size $\varepsilon$ and the mass matrix $M$ during warmup. 

Thus moving from HMC to NUTS further improves exploration efficiency, but at the cost of additional computational complexity arising from adaptive path-building and termination criteria. 
Conveniently, Stan performs these computations under the hood, so this complexity is hidden from the user. 
As with HMC, NUTS requires the log-likelihood to be differentiable; however for ODE models this is not a limitation, as common ODE solvers provide built-in sensitivity calculations that compute parameter derivatives alongside the solution.

\subsection{Revisiting Bayes' rule for estimating parameters in ODE models}
\label{sec:bayesODE}
We are now ready to apply what we have learned about Bayesian modelling and Stan to infer parameter values in ODE models from time series data.
To this end, let us consider a biological system which generates observed data that can be represented as an ODE solution plus noise such that
\begin{equation}
    \tilde y_{i} = y(t_i; \xi) + \epsilon_{i}, 
     \qquad y(0)=y_0,
    \label{eq:ode_plus_noise}
\end{equation}
where {$\xi$} denotes the ODE model parameters, 
$y(t_i;\xi$) is the deterministic solution of the ODE evaluated at time $t_i$, $\tilde y_{i}$ is the $i$th observed data point in the time series, 
$y_0$ is the initial state value,  
and $\epsilon_{i}$ is the corresponding noise term. 
If we, for simplicity, assume that each noise term $\epsilon_{i}$ is independently drawn from a time-invariant normal distribution with mean $\mu=0$ and variance $\sigma^2$, then
\begin{equation}
\epsilon_{i} \sim \mathcal{N}(0,\,\sigma^{2}).
\label{eq:noise_term}
\end{equation}
Combining Eqs.~\eqref{eq:ode_plus_noise} and \eqref{eq:noise_term}, and using the assumed independent zero-mean Gaussian noise model centered on the ODE solution, we arrive at
\begin{equation}
   \tilde y_{i} \sim \mathcal{N} \big( y(t_i ;\xi), \sigma^2 \big).
    \label{eq:ode_stan}
\end{equation}

\noindent With Stan, we can infer the joint posterior distribution of the parameters of interest. 
Common choices for these parameters include ODE model parameters $\xi$, noise parameters such as $\sigma$, and initial conditions $y_0$. Collectively, these are represented by $\theta$ in Bayes’ rule (Eq.~\ref{eq:BayesFormula}). 

\begin{figure}
    \centering
    \includegraphics[width=1\linewidth]{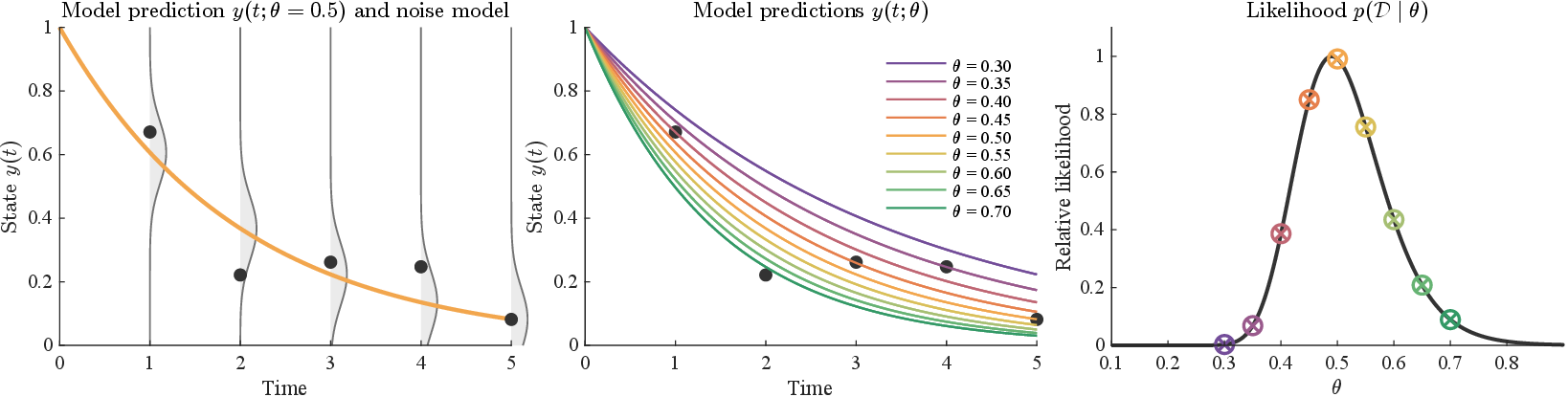}
    \caption{
    {\bf A visual demonstration of how a likelihood function for a model parameter is constructed from an ODE formulation, a noise model, and observed data.}
    Left: The line shows a forward simulation of the ODE model
    \( dy(t)/dt = -\theta\, y(t)\) with initial condition \(y(0)=1\), and 
    \(\theta=0.5\).
    Black points indicate observed data.
    The noise model is fixed as 
    \(\tilde{y}_i = y(t_i;\theta) + \varepsilon_i\), 
    with \(\varepsilon_i \sim \mathcal{N}(0,\sigma^2)\) and \(\sigma=0.1\).
    Shaded ridges indicate the probability density of observing data at each measurement time under the chosen ODE, parameter value \(\theta\), and noise model.
    Middle: The lines show forward simulations of the ODE model \(y(t;\theta)\) 
    for different fixed values of \(\theta\), 
    illustrating how changes in the parameter affect the model’s ability to predict the observed data.
    Right: The likelihood is constructed by evaluating the discrepancy between the observed data $\mathcal D = \{\tilde y_i\}_{i=1}^5$ and the corresponding model predictions \(y(t_i;\theta)\) for each candidate parameter value under the fixed noise model.
    Markers indicate the parameter values explored in the middle panel, 
   following the same colour legend. 
    }
    \label{fig:ode_inf}
\end{figure}

In practice we only require the simplified, proportional version of Bayes' rule (Eq.~\ref{eq:BayesFormula_prop}). As such, running inference in Stan requires three components: 
(i) the observed data, 
(ii) a likelihood function describing how the data are generated by the model, and 
(iii) prior distributions for the parameters being inferred (which default to flat priors if unspecified).
For (ii), it is important to note that the ODE model, its parameters, and the noise model are all embedded in the likelihood, as illustrated in Fig.~\ref{fig:ode_inf}.
Once the full specification (i–iii) is provided, we have all the components necessary to run Stan's sampling algorithm. 
Given the observed time series data $\mathcal D = \{\tilde y_i\}_{i=1}^N$, and with the simplifying assumption that all $N$ data points are conditionally independent given $\theta$, the unnormalised posterior factorises as
\begin{equation}
  p(\theta \mid \mathcal{D}) \propto p(\theta) \prod_{i=1}^{N} p(\tilde y_i \mid \theta).
  \label{eq:thetaprod} 
\end{equation}
By log-transforming both sides we obtain 
\begin{equation}
  \log p(\theta \mid \mathcal{D})
  \propto \log p(\theta) + \sum_{i=1}^{N} \log p(\tilde y_i \mid \theta),
  \label{eq:thetasum}
\end{equation}
where Stan uses this unnormalised log-posterior as its target density (Eq.~\ref{eq:target}). 
With all components (i–iii) specified, Stan runs a specialised NUTS algorithm ``under the hood'' to sample from this target distribution. The gradients of the target log-posterior drive the Hamiltonian dynamics, while its value is used in the accept--reject step.
The resulting samples provide us with an approximation of the joint parameter distribution that can be marginalised to obtain each ODE model parameter.

\subsection{A shallow description of different pooling methods}
\begin{figure}[htbp]
    \centering
    \begin{subfigure}{0.3\textwidth}
        \centering
        \begin{tikzpicture}

´        \node[rectangle, draw, fill=gray!10,
        minimum width=1.4cm, minimum height=1.8cm,
        rounded corners] at (0,-2.4)  {};
        
        \node[circle, draw, minimum size=1.1cm] at (0,0) (q) {$\theta$};
        \node[circle, draw, minimum size=1.1cm, fill=gray!30] at (0,-2.6) (y) {$Y_m$};

        \node at (1.4,-3.2) (t) {$\scriptstyle m=1,\dots,M$};

        \draw[->, thick] (q) -- (y);
        
        \end{tikzpicture}
        \caption{Complete pooling.}\label{fig:comp-pool}
    \end{subfigure}
    \begin{subfigure}{0.3\textwidth}
        \centering
        \begin{tikzpicture}
        \node[rectangle, draw, fill=gray!10,
        minimum width=1.4cm, minimum height=4.0cm,
        rounded corners] at (0,-1.3) {};
      
        \node[circle, draw, minimum size=1.1cm, fill=white] at (0,0) (q) {$\theta_{(m)}$};
        \node[circle, draw, minimum size=1.1cm, fill=gray!30] at (0,-2.6) (y) {$Y_m$};

        \node at (1.4,-3.2) (t) {$\scriptstyle m=1,\dots,M$};

        \draw[->, thick] (q) -- (y);
        \end{tikzpicture}
        \caption{No pooling.}\label{fig:no-pool}
    \end{subfigure}
    \begin{subfigure}{0.3\textwidth}
        \centering
        \begin{tikzpicture}

        \node[rectangle, draw, fill=gray!10,
        minimum width=1.4cm, minimum height=2.8cm,
        rounded corners] at (0,-1.9)  {};

        \node[circle, draw, minimum size=1.1cm] (mu) at (-1.2,0) {$\mu_\theta$};
        \node[circle, draw, minimum size=1.1cm] (sigma) at (1.2,0) {$\sigma_\theta$};
        \node[circle, draw, minimum size=1.1cm, fill=white] (theta) at (0,-1.2) {$\theta_{(m)}$};
        \node[circle, draw, minimum size=1.1cm, fill=gray!30] (y) at (0,-2.6) {$Y_m$};

        \draw[->, thick] (mu) -- (theta);
        \draw[->, thick] (sigma) -- (theta);
        \draw[->, thick] (theta) -- (y);

        \node at (1.4,-3.2) (t) {$\scriptstyle m=1,\dots,M$};

        \end{tikzpicture}
        \caption{One-layer partial pooling.}\label{fig:part-pool}
    \end{subfigure}

    \caption{
    {\bf A visual representation of different pooling strategies in Bayesian models.} 
    Nodes represent random variables and edges indicate conditional dependence. 
    $Y_m$ denotes observed random variables 
    from time series $m$, and $\theta$, $\theta_{(m)}$ are (possibly multivariate) parameters. 
    Plates (shaded rectangles) indicate that the variables inside them are repeated $M$ times. 
    Left: With complete pooling, each $Y_m$ is assumed to have been generated from a common parameter $\theta$.
    Middle: With no pooling, each $Y_m$ is assumed to have been generated from its own parameter $\theta_{(m)}$.
    Right: With partial pooling, each $Y_m$ is assumed to have been generated from its own parameter $\theta_{(m)}$, where all $\theta_{(m)}$ are drawn from a group-level distribution, here parameterised with mean $\mu_\theta$ and variance $\sigma_\theta$. 
    }
    \label{fig:pooling}
\end{figure}
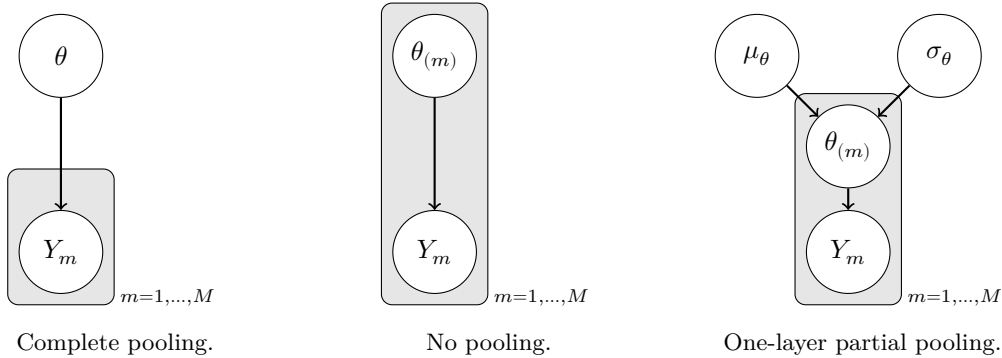

When we collect data from several related sources, we have to decide how much these data should be analysed together versus separately. In other words, we need to decide how to \textit{pool} the data.
In the context of multiple time series (e.g., measurements from different experimental wells, ecological sites, or patients), this question determines how parameter information is shared across the different series. 
In Fig.~\ref{fig:pooling} we illustrate three standard pooling strategies: complete, no, and partial pooling.

If we assume that all observed data were generated by the same biological process with a single (possibly multivariate) parameter, then we use complete pooling to infer a shared posterior distribution over $\theta$.
In this case, evaluating the target density requires extending the index $i$ in Eq.~\eqref{eq:ode_plus_noise} to a tuple $(i,j)$, jointly indexing observations across time series and measurement points, so that Eq.~\eqref{eq:thetasum} becomes a double sum over $i$ and $j$. 
If, on the other hand, we assume that each series is better described by its own parameter, then we use a no pooling approach to infer a separate posterior distribution over $\theta_{(m)}$ for each series $m$ by applying Eq.~\eqref{eq:thetasum} independently.
Between these two extremes lies partial pooling. In this case, each time series has its own individual-level parameter $\theta_{(m)}$ drawn from a higher, group-level distribution that is characterised by some hyperparameters, such as a mean $\mu_\theta$ and standard deviation $\sigma_\theta$ in the case of a normal distribution. 
These hyperparameters may, in turn, be drawn from even higher-level distributions. 
As such, partial pooling allows us to build hierarchical structures that capture relationships across groups and individuals. 
The choice of pooling strategy should be made on a case-by-case basis, guided by the data and the research question at hand.
A natural first option is often complete pooling, as it involves the fewest parameters, is generally easier to infer than partial pooling, and is more robust for predicting unseen data than no pooling \cite{gelman1995bayesian}.

\subsection{Diagnosing sampling behaviour}\label{sec:diagnostics}
Having discussed how to use Stan for Bayesian inference, we now turn to an equally important task: assessing whether these inferences are actually valid. 
To avoid bias, it is important to ensure that the sampling method has adequately explored the parameter space $\Theta$ before drawing any conclusions about the target
density $\pi(\theta)$. 
In practice, inference is based on running multiple sampling trajectories, called chains, rather than a single one. These chains are initialised at different points in $\Theta$, and will explore the posterior space along different stochastic paths. 
The posterior is then approximated by aggregating the samples from all chains after discarding their initial burn-in phases. 
Analysing behaviour within and between chains helps us reveal sampling pathologies. 
To this end, there exist multiple diagnostic metrics that help assess the quality of the generated samples and detect potential pathological behaviour in the sampling process.
In this section, we provide an intuitive overview of three commonly used diagnostic metrics, all of which are computed and reported by default in Stan. 

\begin{figure}[h!]
\centering
\includegraphics[width=\textwidth]{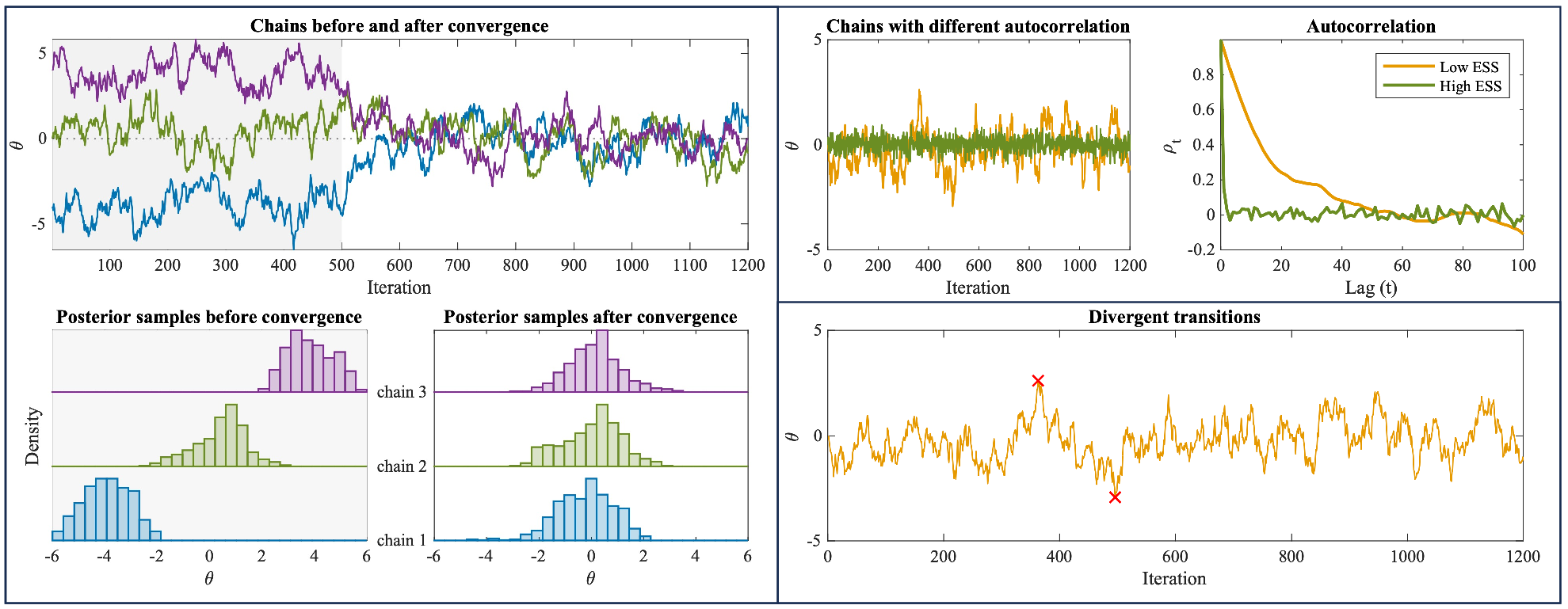}
\caption{
{\bf Illustrative example of sampling diagnostics assessed with $\widehat{R}$ (left), effective sample size (top right), and divergent transitions (bottom right).}
Colours are matched across panels to indicate the same sampling chain.
Left: Three sampling chains and their corresponding  density estimates are shown before convergence (when $\widehat{R} > 1$) and after convergence (when $\widehat{R} \approx 1$).
Top right: Two sampling chains are compared: one with high effective sample size (ESS) and one with low ESS, computed using single-chain ESS. 
The panels show how the chains evolve over sampling iterations and how their autocorrelation decays with increasing lag. 
Bottom right: A sampling chain in which divergent transitions are marked with red crosses.
}
\label{fig:diagnosis}
\end{figure}

\paragraph{$\widehat{R}$ statistic} 
The $\widehat{R}$ statistic is designed to assess whether all Markov chains have converged to the same stationary target distribution, i.e., a distribution that no longer depends on the chains' initial positions \cite{gelman1992inference}. 
If yes, we can intuit that the chains would mix at stationarity; if no, the chains would remain distinguishable from one another, as illustrated in Fig.~\ref{fig:diagnosis}.
The $\widehat{R}$ statistic captures this idea quantitatively through its definition,
\begin{equation} \label{eq:rhat}
\widehat{R}
= \sqrt{\frac{\widehat{\mathrm{var}}^{+}(\theta)}{W}}
= \sqrt{\frac{\tfrac{N-1}{N}W + \tfrac{1}{N}B}{W}},
\end{equation}
where $W$ is obtained by computing the sample variance within each chain and averaging these across chains, and $B$ is obtained by first computing the mean of each chain and then computing the variance of these means across chains, scaled by the number of samples $N$. Further, 
$\widehat{\mathrm{var}}^{+}(\theta)$ is a computational estimate of the marginal posterior variance $\mathrm{Var}_{\pi}(\theta)$.
Thus, if $\theta$ is multivariate, $\widehat{R}$ is computed separately for each component of $\theta$.
As the chains explore the target distribution in $\Theta$, 
the within-chain variance $W$ approaches the marginal posterior variance $\mathrm{Var}_{\pi}(\theta)$. 
On the other hand, during convergence, $\widehat{\mathrm{var}}^{+}(\theta)$ decreases toward this same value as between-chain differences vanish, 
so that $\widehat{R}$ approaches one.
As such, $\widehat{R}\approx 1$ indicates mixing and convergence to the stationary target, while $\widehat{R}>1$ indicates that the chains have not converged. 
Stan reports a modified version of Eq.~\eqref{eq:rhat} called the rank-normalised $\widehat{R}$ which accounts for the full empirical distribution when assessing convergence to the stationary target distribution, addressing variance-related limitations, such as when dealing with heavy-tailed or infinite variance distributions \cite{vehtari2021rank}.

\paragraph{Effective sample size} 
An ideal sampling method would produce completely independent samples from the posterior distribution. However, MCMC-based samplers like HMC are built on Markov chains which, by design, result in correlation between subsequent samples. This dependence reduces the amount of  information captured per sample. The effective sample size (ESS) provides a measure of the efficiency of the sampler after adjusting for such autocorrelation. It is defined as
\begin{equation}
    \textrm{ESS} = \frac{N}{\sum_{t=-\infty}^\infty\rho_t} = \frac{N}{1+2\sum_{t=1}^\infty \rho_t},
\end{equation}
where $N$ denotes the number of samples per chain and $\rho_t$ is the within-chain autocorrelation at lag $t$, i.e.\ the correlation between samples separated by $t$ iterations (Fig.~\ref{fig:diagnosis}). In practice, the autocorrelation cannot be computed exactly, and is instead estimated based on the samples. 
From an inference perspective, the ESS must be large enough to ensure that estimates obtained from different simulation runs are consistent with each other.

In Stan, two measures of ESS are available. 
The first, referred to as bulk-ESS, measures how effectively the samples can estimate quantities in the bulk of the distribution (e.g., the mean or median).
The second, referred to as tail-ESS, measures how effectively the samples can estimate quantities in the tails of the target distribution (e.g., extreme quantiles).
By default, Stan computes effective sample sizes by combining information across all sampling chains, accounting for both within-chain autocorrelation and between-chain variability.

\paragraph{Divergences}

Posterior geometries can be complex and difficult to explore; a classical example of this is the funnel-shape typically induced by hierarchical models \cite{betancourt2015hamiltonian}. 
Difficult posterior geometries often arise in regions of the target density with high curvature, where Hamiltonian trajectories change rapidly and are therefore challenging to integrate accurately with fixed step sizes. 
This is problematic because HMC methods rely on discrete (sympletic) integrators in practice, and thus numerical integration errors can accumulate over time. 
When this happens, the Hamiltonian energy is no longer well conserved, 
resulting in a difference in energy between at $\theta'$ and $\theta$,
\begin{equation}
    |\mathcal{H}(\theta',r') - \mathcal{H}(\theta,r)|  > \delta.
\end{equation}
When this difference is large, the behaviour is flagged as a divergent transition in Stan, indicating that the Hamiltonian trajectory underlying the transition was numerically unstable. 
By default, this threshold is set to $\delta = 10^3$ in Stan.
As a result, inferences drawn from chains exhibiting divergent transitions should be treated with caution, since the resulting samples may not reliably represent the target density. 
Divergence issues can often be resolved by adjusting the model or simulation hyperparameters, or in some cases, by model reparameterisation \cite{betancourt2015hamiltonian}. 
In diagnostic plots, Stan marks divergent transitions, similar to what is done in Fig.~\ref{fig:diagnosis}.
%

\enlargethispage{1\baselineskip}
\subsection{Evaluating model performance}
When constructing a model, we want it to fit the observed data well, but we also want it to perform well on unseen data.
In Bayesian frameworks, model performance is commonly assessed using the log predictive density (lpd) and the expected log predictive density (elpd).
The former evaluates model agreement with the observed data, and the latter with unseen data.
We describe both quantities below and, for simplicity, illustrate them for a single time series.

\paragraph{Log predictive density (lpd)}
As seen in Eq.~\eqref{eq:thetasum}, the posterior is constructed from a sum of pointwise log-likelihood contributions evaluated at the observed data $\mathcal D = \{\tilde y_i\}_{i=1}^N$. 
To assess how well the fitted model predicts these observations, we let $Y_i^*$ denote a future observation at time $t_i$. 
The posterior predictive distribution of $Y_i^*$ is obtained by integrating over all possible parameter values weighted by their posterior probability,
\begin{equation}
p( Y_i^* \mid \mathcal D) 
=
\int p(Y_i^* \mid \theta)\, p(\theta \mid \mathcal D)\, d\theta.
\end{equation}
Evaluating this density at the observed value $\tilde y_i$, taking the logarithm, and summing over all observations gives the log predictive density (lpd),
\begin{equation}
\mathrm{lpd}
=
\sum_{i=1}^{N}
\log p(\tilde y_i \mid \mathcal D),
\label{eq:lpd}
\end{equation}
which gives a rough measure of how probable the data is under the model. In practice, this quantity is approximated using posterior predictive draws generated from the posterior samples produced by Stan.

\paragraph{Expected log predictive density (elpd)}
The lpd in Eq.~\eqref{eq:lpd} evaluates model agreement on the same data used for inference and is therefore optimistically biased; relying solely on the lpd tends to overestimate model performance. 
To assess out-of-sample performance, we instead wish to consider new data points $y_i^*$ drawn from the true (but typically unknown) data-generating distribution $p_t(y_i^*)$.
Taking the expected value of the lpd with respect to this distribution and summing over all new data points leads to the elpd,
\begin{equation}
\mathrm{elpd}
=
\sum_{i=1}^{N}
\int
p_t(y_i^*)
\log p(y_i^* \mid \mathcal D)
\, dy_i^*.
\label{eq:elpd}
\end{equation} 
In practice, Eq.~\eqref{eq:elpd} cannot be computed exactly, as the true data-generating distribution is unknown. 
Instead, it is approximated using methods such as leave-one-out cross-validation (LOO-CV) or the Watanabe--Akaike information criterion (WAIC), also called the widely applicable information criterion \cite{vehtari2017practical}. 
For models fitted in Stan, the \textsc{loo} package provides an efficient approximation of the elpd, yielding the estimate $\widehat{\mathrm{elpd}}_{\mathrm{LOO}}$ \cite{vehtari2015loo}. 

As part of a model selection procedure, the elpd can be used to compare the predictive performance of different models, with larger values indicating better predictive accuracy. 
When comparing two models, the difference between their estimated elpd values is then evaluated relative to the standard error of this difference, which reflects the variability of the pointwise predictive contributions across observations. 
The elpd difference (\texttt{elpd\_diff}) and its standard error (\texttt{se\_diff}) can be computed using the \textsc{loo} package and, as a rough guideline,  absolute differences larger than about twice the standard error may indicate meaningful differences in predictive performance \cite{vehtari2017practical}.

\enlargethispage{1\baselineskip}
\section{Implementing ODE models in Stan}
\subsection{A toy model of bacteria population dynamics}
\label{sec:exsyst}
To demonstrate how to implement and analyse ODE models in Stan, we consider a simulated bacterial co-culture experiment. 
Two subpopulations are grown together and compete for space in $W=6$ replicated wells.
Subpopulation sizes are recorded every 4 hours over a 48-hour period in each well, yielding longitudinal time-series data exhibiting logistic-type growth. 
Accordingly, we model the subpopulation dynamics using the following competition model,
\newpage
\begin{linenomath}
\begin{subequations}
\label{eq:toymodel}
\begin{align}
\frac{dy_1(t)}{dt} & = r_1 y_1(t)\!\left(1-\frac{y_1(t)+y_2(t)}{K}\right), \label{eq:toymodel_a}\\
\frac{dy_2(t)}{dt} & = r_2 y_2(t)\!\left(1-\frac{y_1(t)+y_2(t)}{K}\right), \label{eq:toymodel_b}
\end{align}
\end{subequations}
\end{linenomath}
\noindent with unknown initial conditions
\begin{equation}
y_1(0)=y_{1,0}, \qquad y_2(0)=y_{2,0}.
\label{eq:toymodel_ic}
\end{equation}
\noindent Here, $y_s(t)$ denotes the size of subpopulation $s \in \{1,2\}$ at time $t$,  $r_1$ and $r_2$ the corresponding intrinsic growth rates, and $K$ the shared carrying capacity. 
At observation times $t_i$ we assume the measurement model
\begin{linenomath}
\begin{subequations}
\label{eq:toymodel_noise}
\begin{align}
\tilde y_{1,i} &= y_1(t_i) + \epsilon_{1,i},  \qquad \epsilon_{1,i} \sim \mathcal{N}(0,\sigma^2), \label{eq:obs_y1}\\
\tilde y_{2,i} &= y_2(t_i) + \epsilon_{2,i}, \qquad  \epsilon_{2,i} \sim \mathcal{N}(0,\sigma^2), \label{eq:obs_y2}
\end{align}
\end{subequations}
\end{linenomath}
where $\tilde y_{s,i}$ denotes the noisy measurement of subpopulation $s$ at time $t_i$, and all noise terms are assumed to be independently sampled from a normal distribution with mean $0$ and variance $\sigma^2$.
Our task at hand is to use Stan to infer the unknown parameters from simulated time-series data generated by Eqs.~\eqref{eq:toymodel}--\eqref{eq:toymodel_noise}. 
\noindent Thus, the parameters to infer are
\begin{equation}
\theta = \big( r_1, r_2, K, y_{1,0}, y_{2,0}, \sigma \big),
\label{eq:theta_toy}
\end{equation}
\noindent with $r_1, r_2, K, \sigma > 0$ and $y_{1,0}, y_{2,0} \ge 0$, to ensure biological and physical realism. 
Time is measured in hours (h), so that growth rates have units of h$^{-1}$, and population sizes, and thus also the carry capacities and noise level, are measured in units of $10^6$ cells per well.

\subsection{Writing a Stan file block-by-block}
\label{sec:write} 
Stan files are built from blocks that structure and compartmentalise the information needed for parameter inference. 
Seven block types are available, and when used, they must appear in a specific order. 
Only the \texttt{model} block is strictly required, but typical Stan files also include the \texttt{data} and \texttt{parameters} blocks. 
The remaining blocks, which are included when needed, are labeled \texttt{functions}, \texttt{transformed data}, \texttt{transformed parameters}, and \texttt{generated quantities}.
In the required top-to-bottom order, the blocks used to implement the bacterial toy model with complete pooling are outlined below and together form the Stan file \texttt{toymodel.stan}. This file is available in the Supplementary Material (SM2), together with the corresponding R code for generating synthetic data, interfacing with Stan, and inspecting inference results and diagnostics.

\enlargethispage{1\baselineskip}
\paragraph{The \texttt{functions} block.}
We define the right-hand side of the logistic growth equation in the \texttt{functions} block.
The function returns the time derivative of the state vector at time \texttt{t}. 
The ODE solver repeatedly evaluates this function for different parameter values \texttt{p} while numerically integrating the system at the requested time points. 

\begin{lstlisting}[caption=The functions block., label={lst:listing}, style=stan]
functions {
  vector logistic_ode(real t, vector y, vector p) {
    vector[2] dydt;
    real r1 = p[1];
    real r2 = p[2];
    real K  = p[3];
    
    // ---- ODE model ----
    dydt[1] = r1 * y[1] * (1 - (y[1] + y[2]) / K);
    dydt[2] = r2 * y[2] * (1 - (y[1] + y[2]) / K);

    return dydt;
  }
}
\end{lstlisting}

\newpage

\paragraph{The \texttt{data} block.}
Data are passed to Stan from an external interface and must be explicitly declared in the \texttt{data} block using built-in types such as \texttt{int} or \texttt{real}, or container types such as \texttt{array} and \texttt{matrix} variables constructed from these. 
Declarations may include upper and lower bounds that constrain the allowed range of the input data and provide a simple error-check.
For the two-species competition model, we pass the following data to Stan:
an integer scalar \texttt{no\_ts} denoting the number of observation times;
an integer scalar \texttt{no\_wells} specifying the number of experimental wells;
a real-valued scalar \texttt{t0} specifying the initial time;
a real-valued array \texttt{ts} of length \texttt{no\_ts} containing the observation times;
a real-valued array \texttt{y\_tilde} of length \texttt{no\_wells}, containing matrices of size \texttt{2}\,$\times$\,\texttt{no\_ts}, where each matrix stores the observed population sizes for the two subpopulations across all time points in a given well;
an integer scalar \texttt{no\_ts\_gen} specifying the number of time points at which predictions are generated;
and a real-valued array \texttt{ts\_gen} of length \texttt{no\_ts\_gen} containing the prediction time grid.
In \texttt{y\_tilde}, the element \texttt{y\_tilde[j][s,i]} represents the measurement from well $j$ and subpopulation $s \in \{1,2\}$, at observation time index $i$.
\begin{lstlisting}[caption=The data block., label={lst:listing}, style=stan]
data {
  int<lower=1> no_ts;                           // Number of observation time points
  int<lower=1> no_wells;                        // Number of wells
  real t0;                                      // Initial time
  array[no_ts] real<lower=t0> ts;               // Observation times
  array[no_wells] matrix[2, no_ts] y_tilde;     // Observed data

  int<lower=1> no_ts_gen;                       // Number of prediction time points
  array[no_ts_gen] real ts_gen;                 // Prediction times
}
\end{lstlisting}

\paragraph{The  \texttt{transformed data} block.} 
Although we do not make use of this block in the toy model example, we describe it here for completeness. 
At runtime, code in the \texttt{data} and \texttt{transformed data} blocks is evaluated  once per chain, whereas code in other blocks may be evaluated repeatedly during sampling. 
As indicated by its name, variables in this block may depend on those declared in the \texttt{data} block or may also be constants required by the model. 
Placing quantities that remain constant throughout the chain in the \texttt{transformed data} block therefore avoids unnecessary recomputation.
\begin{lstlisting}[caption=The transformed data block., label={lst:listing}, style=stan]
transformed data {
  //transformed data here
}
\end{lstlisting}

\paragraph{The  \texttt{parameters} block.} 
All unknown quantities to be inferred are declared in the \texttt{parameters} block.
For the logistic growth model, these are the intrinsic growth rates $r_1,r_2$, the carrying capacity $K$, the initial population size $y_{1,0}, y_{2,0}$, and the observation noise parameter $\sigma$.
In the Stan implementation, the growth parameters are collected in the vector \texttt{p}. 
When declaring parameters, lower and upper bounds may optionally be imposed, where Stan enforces these constraints during sampling. 
In this example, positivity constraints are used to ensure biological feasibility, restricting the parameters to positive values.

\begin{lstlisting}[caption=The parameters block., label={lst:listing}, style=stan]
parameters {
  vector<lower=0>[3] p;                         // p[1]=r1, p[2]=r2, p[3]=K
  real<lower=0> sigma;                          // Observation noise
  vector<lower=0>[2] y0;                        // Initial values for both species
}
\end{lstlisting}

\enlargethispage{1\baselineskip}
\paragraph{The \texttt{transformed parameters} block.}
The \texttt{transformed parameters} block is used to compute quantities that depend on the model parameters and are required repeatedly during sampling. 
In our implementation, the ODE system is solved at the observation times \texttt{ts} for each sampled parameter vector. 
The numerical solution returned by the ODE solver is first stored in \texttt{y\_state}, which contains the deterministic state vectors of both subpopulations at each time point.
For convenience in specifying the likelihood, the deterministic trajectories are rearranged into the matrix \texttt{y}, whose rows correspond to subpopulations and whose columns correspond to observation times.
The matrix \texttt{y} therefore represents the deterministic mean dynamics that enter the Gaussian measurement model in the \texttt{model} block.

\begin{lstlisting}[caption=The transformed parameters block., label={lst:listing}, style=stan]
transformed parameters {
  vector[2] y0_state;                           // Initial state vector
  array[no_ts] vector[2] y_state;               // Deterministic ODE solution (solver output)
  matrix[2, no_ts] y;                           // Deterministic trajectories (for likelihood)

  // ---- Solve ODE system at observation times ----
  y0_state = y0;
  y_state = ode_rk45(logistic_ode, y0_state, t0, ts, p);

  // ---- Convert solver output to 2 x no_ts matrix ----
  for (i in 1:no_ts) {
    y[1, i] = y_state[i][1];
    y[2, i] = y_state[i][2];
  }
}
\end{lstlisting}

\vspace{-.3cm}
\paragraph{The \texttt{model} block.}
In the \texttt{model} block, we assign prior distributions to the parameters declared in the \texttt{parameters} block and specify the likelihood induced by the two-species competition ODE system and the observation model.
Because the coupled nonlinear system does not admit a closed-form solution, it is solved numerically using Stan’s built-in ODE solver \texttt{ode\_rk45}. 
The ODE solution is computed in the \texttt{transformed parameters} block, where it is stored in the variable \texttt{y}, representing the deterministic mean trajectories of both subpopulations at the observation times.
In the \texttt{model} block, we then condition on this precomputed trajectory
and specify a Gaussian measurement model with  standard deviation \texttt{sigma}.

\begin{lstlisting}[caption=The model block., label={lst:listing}, style=stan]
model {
  // ---- Priors ----
  p[1] ~ normal(0.2, 0.1);                      // Growth rate r1
  p[2] ~ normal(0.2, 0.1);                      // Growth rate r2
  p[3] ~ normal(5, 1);                          // Shared carrying capacity K
  sigma ~ normal(0, 0.5);                       // Noise standard deviation
  y0 ~ normal(1, 0.5);                          // Initial values

  // ---- Likelihood (elementwise) ----
  for (j in 1:no_wells) {
    to_vector(y_tilde[j]) ~ normal(to_vector(y), sigma); 
  }
}
\end{lstlisting}

\vspace{-.3cm}
\paragraph{The \texttt{generated quantities} block.}
After sampling from the joint posterior distribution of the parameters, each posterior draw can be used to generate predicted model dynamics. 
In Stan, such posterior predictive simulations are implemented in the \texttt{generated quantities} block. 
For each sampled parameter vector, the ODE system is solved on the specified prediction time grid \texttt{ts\_gen} to obtain the deterministic mean trajectories of both species. 
These noise-free trajectories, stored in \texttt{y\_mean}, are useful for mechanistic interpretation, as they reflect uncertainty in the inferred model parameters while excluding measurement error.
To obtain predictions on the observation scale, Gaussian noise with standard deviation $\sigma$ is added to the deterministic solution, yielding 
samples from the posterior predictive distribution 
stored in \texttt{y\_pred}.
These posterior predictive simulations can be used to assess model performance by visually comparing predicted and observed data, or by computing quantitative discrepancy measures.
Although posterior predictions can also be generated outside Stan, implementing them within a Stan file ensures that they are automatically computed for every posterior draw and stored together with the sampled parameters.

\begin{lstlisting}[caption=The generated quantities block., label={lst:listing}, style=stan]
generated quantities {
  // ---- Predictions at time points ts_gen ----
  array[no_ts_gen] vector[2] y_state_pred;      // Deterministic ODE solution (solver output)
  matrix[2, no_ts_gen] y_mean;                  // Noise-free trajectory 
  matrix[2, no_ts_gen] y_pred;                  // Posterior predictive draws (with noise)

  vector[2] y0_state_gen;
  y0_state_gen = y0;

  // ---- Solve ODE at prediction times ----
  y_state_pred = ode_rk45(logistic_ode, y0_state_gen, t0, ts_gen, p);

  // ---- Store noise-free mean + generate noisy observations ----
  for (i in 1:no_ts_gen) {
    y_mean[1, i] = y_state_pred[i][1];         // Noise-free ODE solution
    y_mean[2, i] = y_state_pred[i][2];                 

    y_pred[1, i] = y_mean[1, i] + normal_rng(0, sigma); // Noisy prediction
    y_pred[2, i] = y_mean[2, i] + normal_rng(0, sigma); 
  }
}
\end{lstlisting}

\enlargethispage{1\baselineskip}

\subsection{Inspecting Stan outputs}

\subsubsection{Diagnostics}
Once a Stan model has been fitted to data, the next step is to investigate the sampling output. 
Any issues encountered during inference will produce a warning message immediately after the sampling routine has finished. 
Such issues include high $\widehat{R}$ values, low ESS values, and the presence of divergences, as discussed in Section \ref{sec:diagnostics}. 
Stan provides built-in functions to summarise posterior inference, including key diagnostic metrics (Fig.~\ref{fig:stanoutput}a). Users can specify which parameters and posterior quantiles to report.
For the toy model example, we consider the six parameters in Eq.~\eqref{eq:theta_toy}, and inspect the median and the 5\% and 95\% posterior quantiles.
This output provides a starting point for post-sampling evaluation; parameter quantiles provide a quick sanity check, while per-parameter $\widehat{R}$ and ESS values help identify parameters for which the sampling chains exhibit poor mixing or insufficient effective sample size.

\subsubsection{Posterior predictive checks}
Given that the sampling output looks satisfactory, the next logical question is: how well does the model fit the data?
%
To answer this question, draws from the posterior predictive distribution can be used to visualise model predictions and quantitatively assess agreement with the observed data.
In our toy model example, posterior predictive draws are stored in the Stan output variable \texttt{y\_pred} and are plotted against observed data in Fig.~\ref{fig:stanoutput}b.
The figure shows predictive uncertainty arising from both parameter uncertainty and observational noise.
This visual form of model checking helps us identify potential model misspecification. For example, here we see that the logistic growth model is consistent with the observed bacterial growth data.
This is trivial in the toy model case, since the data were simulated from the same ODE and noise model used for inference. 
In real-world applications, however, the data-generating process is unknown, and posterior predictive checks help assess whether a model adequately captures the observed system dynamics. 
For a broader discussion of iterative Bayesian model development and evaluation, see the Bayesian workflow framework proposed by Gelman and colleagues \cite{gelman2020bayesianworkflow}.

\subsubsection{Parameter distributions}
\begin{figure}[h!]
\vspace*{\fill}
    \includegraphics[width=\linewidth]{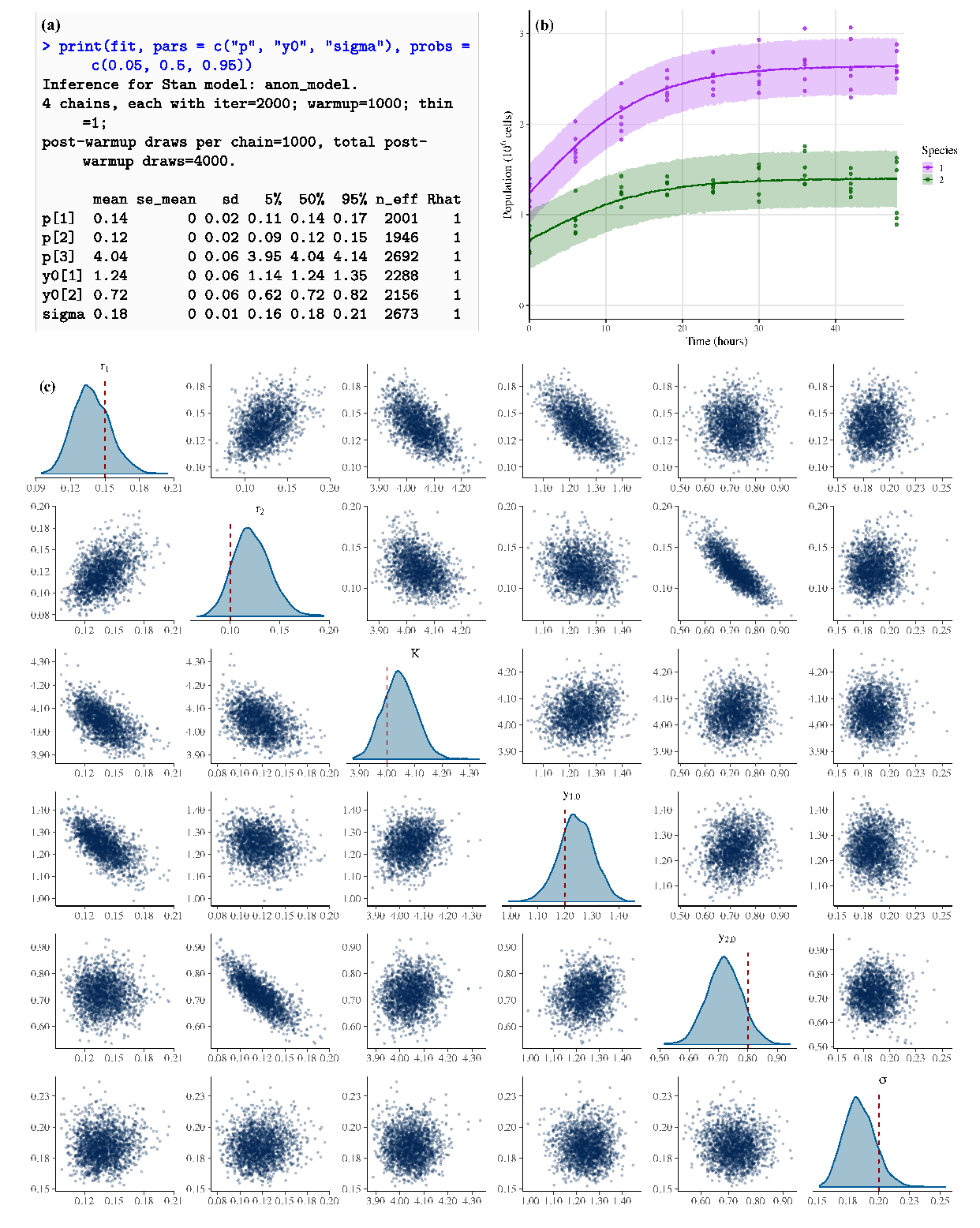}
    \caption{{\bf Stan output for the toy model illustrating posterior summaries and diagnostics, posterior predictive checks, and posterior parameter distributions.}
    (a) Summary statistics of the inferred parameters, including  posterior means, standard deviations, selected quantiles, and sampling diagnostics.
    (b) Posterior predictive distributions plotted against observed 
    (c) Diagonal plots show marginal posterior parameter distributions with the true parameter values (vertical lines) overlaid. 
    Off-diagonal plots show bivariate posterior parameter distributions.}
    \label{fig:stanoutput}
    \vspace*{\fill}
\end{figure}

To gain a deeper understanding of the inference results, we can examine the full posterior distributions of the parameters  rather than relying solely on summary statistics.
However, visualising the full joint posterior distribution is only feasible in low-dimensional parameter spaces, which are uncommon in real-world biological applications.
Instead, we can examine marginal and pairwise posterior distributions to gain insight into the structure of the joint posterior.
Empirical posterior distributions are commonly visualised using either a histogram or a smooth density estimate, the latter of which is used in the diagonal plots in Fig.~\ref{fig:stanoutput}c. 
Bivariate visualisations depict the posterior density for pairwise parameter combinations, where areas of high density are indicated by the colour or intensity at that point.
In the figure, the off-diagonal panels show bivariate posterior densities for parameter pairs.
Such pair plots make it easy to identify correlations between parameters and can highlight regions of the posterior associated with sampling issues, such as divergences.

\enlargethispage{1\baselineskip}

\section{Examples with real data}
In this section we show how real-world dynamical systems can be implemented in Stan using two examples borrowed from previous mathematical modelling studies: one on coral coverage in the Great Barrier Reef, and one on biomarker dynamics in prostate cancer patients. 
The code and data used for both examples are available on GitHub\footnote{See \url{https://github.com/jodie-c/stan_tutorial}}.

\newpage 

\newpage

\subsection{Coral bleaching in the Great Barrier Reef} 
Coral bleaching is primarily driven by global warming and stands as one of the major threats to coral reefs \cite{hughes2003climate}.
Given the significant ecological value of coral reefs, the severity of recent bleaching events, and the complex nature of potential recovery, a deeper understanding of coral reef dynamics is crucial. 
The Australian Institute of Marine Science (AIMS) has instigated a comprehensive and long-running effort to monitor the condition of the Great Barrier Reef, known as the Long-Term Monitoring Programme (LTMP) \cite{aims}. 
The collected data provide longitudinal observations of multiple sites across different reefs, enabling the monitoring of bleaching events and, in many cases, subsequent recovery.
To model coral cover dynamics, we here use the AIMS LTMP dataset together with an ODE model developed by Brown et al.\cite{brown2025coral}.

\begin{figure}[p]
\centering
\vspace*{\fill}
    \includegraphics[width=\linewidth]{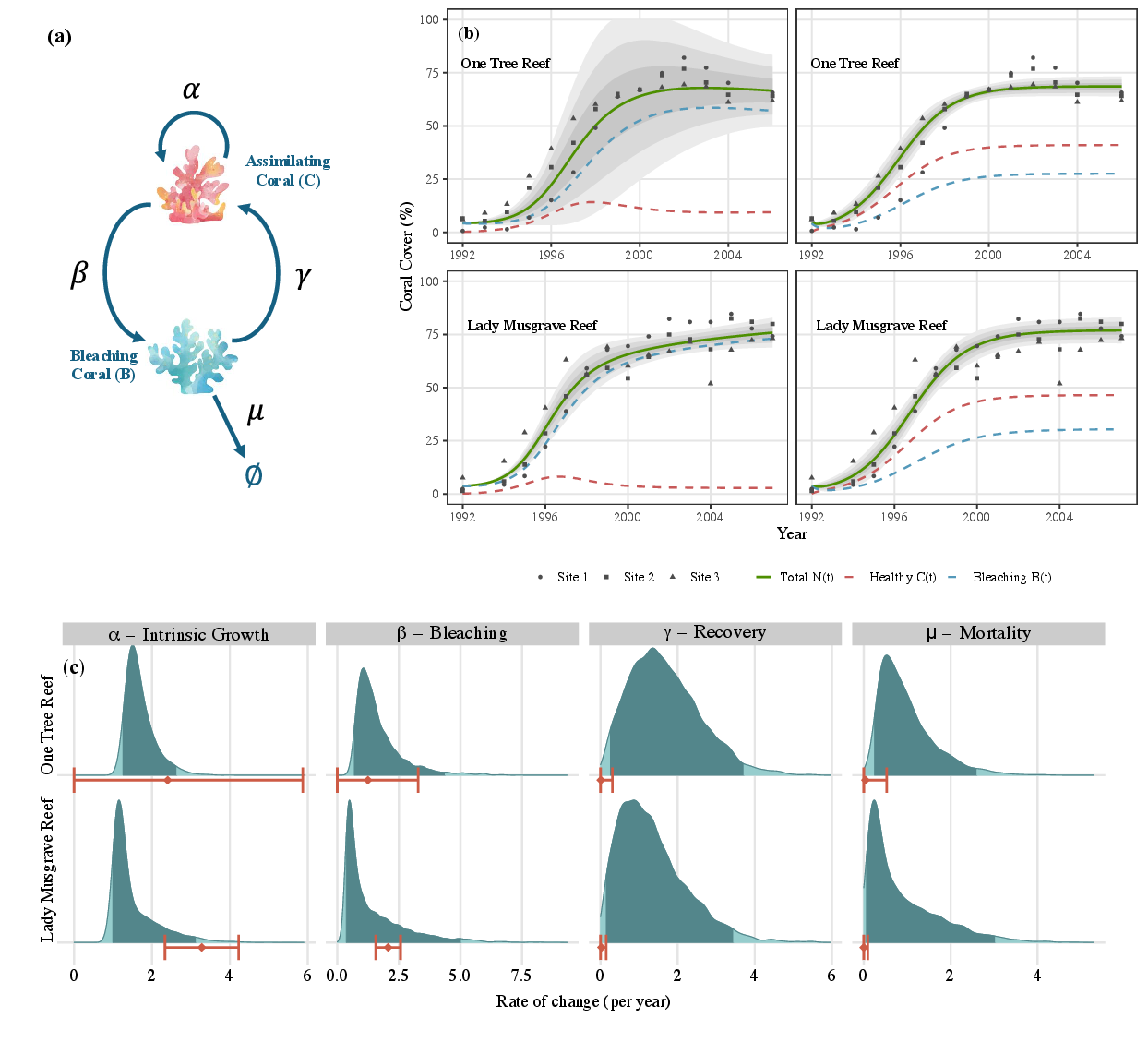}
    \caption{\textbf{System schematic, posterior predictive distributions, and parameter inference for the coral bleaching model.}
    (a) Healthy coral ($C$) transitions to the bleached state at rate $\beta$ and regenerates at the intrinsic growth rate $\alpha$. Bleached coral ($B$) either assimilates (returns to the healthy state) at rate $\gamma$ or dies at rate $\mu$. 
    (b) Total coral cover predictions. The left panels show predictions from the frequentist approach, with shaded regions indicating one, two, and three standard deviations obtained via bootstrap refitting.  
    The right panels show the corresponding Bayesian predictions obtained from posterior predictive simulations, with shaded regions indicating one, two, and three standard deviations of the posterior predictive distribution.
    (c) 
    Overlaid parameter estimates for frequentist and Bayesian methods. For the frequentist approach, the maximum-likelihood estimate and 95\% confidence interval are shown along the axis. For the Bayesian approach, the full posterior distribution is shown, with the 95\% credible interval highlighted by the darker region.}
    \label{fig:coral-reef-example}
  \vspace*{\fill}
\end{figure}

For each site within a reef, we let $C(t) \geq 0$  and $B(t) \geq 0 $ represent the fraction of coral cover in the assimilating (i.e., healthy) and bleaching states, respectively, at time $t$. 
The system dynamics are described by
\begin{linenomath}
\begin{subequations}\label{eq:coral-model}
\begin{align}
    \frac{dC(t)}{dt} &= \alpha C(t)\Big(1-(C(t)+B(t))\Big)-\beta C(t) + \gamma B(t), \\
    \frac{dB(t)}{dt} &= \beta C(t) - \gamma B(t) - \mu B(t),
\end{align}
\end{subequations}
\end{linenomath}
with initial conditions
\begin{equation}
C(0) = C_0, \qquad B(0) = B_0.
\end{equation}
Here, $\alpha$ is the intrinsic growth rate of assimilating coral, $\beta$ is the rate at which coral in the assimilating state transitions to the bleaching state, $\gamma$ is the recovery rate at which bleaching coral returns to the assimilating state, and $\mu$ is the mortality rate of coral in the actively bleaching state. This process is schematically illustrated in Fig.~\ref{fig:coral-reef-example}a. Note that all parameters must be strictly positive. 
\\

\noindent While the ODE has two dependent variables, data are only available in terms of the total coral coverage, i.e., the sum of healthy and bleaching coral, for each site $j$ at observation times $t_i$. 
Thus, assuming normally distributed noise, the observation model becomes
\begin{equation}
    \tilde y_{j,i} = y_{j,i}(t_i;\xi_r) + \epsilon_{j,i}, \qquad \epsilon_{j,i} \sim \mathcal{N}(0,\sigma_r^{2}),
\end{equation}
for reef-specific ODE parameters $\xi_r$ and observation noise standard deviations $\sigma_r$, with total coral coverage given by 
\begin{equation}
     y_{j,i}(t_i;\xi_r) = C_j(t_i) + B_j(t_i).
\end{equation} 
We further pool all sites $j$ belonging to the same reef $r$ using a reef-level complete pooling strategy in order to infer the reef-specific parameters $\theta_r = (\alpha_r, \beta_r, \gamma_r, \mu_r, \sigma_r)$ given fixed initial conditions that are determined from the data. 
The inferred parameter values and the resulting dynamics are compared with those obtained using a frequentist approach \cite{brown2025coral} in Fig.~\ref{fig:coral-reef-example} for two reefs: the Lady Musgrave Reef and the One Tree Reef.
Results for an additional seven reefs are provided in the Supplementary Material (SM3).

Both the frequentist and Bayesian approaches produce similar mean predictions of total coral cover over time, as shown in Fig.~\ref{fig:coral-reef-example}b.
Shaded regions indicate prediction uncertainty: in the frequentist analysis these are obtained via bootstrap refitting of the model, whereas in the Bayesian case they arise from posterior predictive simulations generated in Stan.
Notably, the frequentist prediction interval for One Tree Reef is substantially wider than that for Lady Musgrave Reef, consistent with the broader confidence intervals observed for the corresponding parameter estimates (Fig.~\ref{fig:coral-reef-example}c). 
These confidence intervals are overlaid with the corresponding Bayesian posterior distributions, where the 95\% credible intervals are highlighted.
Although both intervals summarise parameter uncertainty, they have different interpretations. 
A 95\% confidence interval represents a range that would contain the true parameter value in 95\% of repeated samples. 
In contrast, a 95\% credible interval represents the range of parameter values containing 95\% of the posterior probability mass given the model and observed data. 
The latter therefore provides a probabilistic description of parameter uncertainty through the full posterior distribution, allowing the relative plausibility of different parameter values to be assessed.
We further note that the inferred dynamics for healthy and bleaching coral when considered separately are not consistent between the frequentist and Bayesian approaches, suggesting possible identifiability issues that lie beyond the scope of this study.

\newpage

\enlargethispage{1\baselineskip}
\subsection{Biomarker dynamics in prostate cancer patients}
\begin{figure}[H]
\centering
    \includegraphics[width=\linewidth]{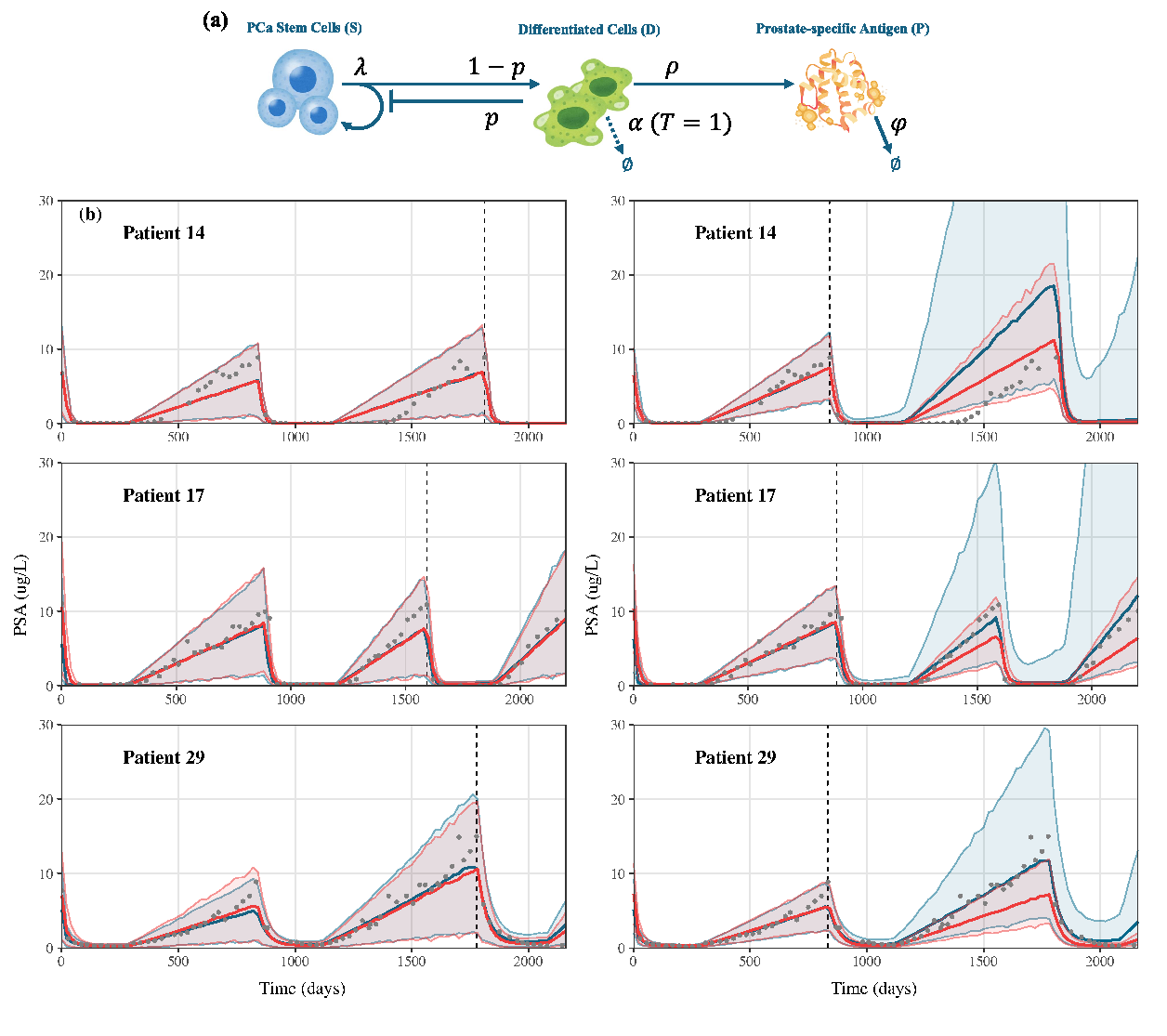}
    \caption{\textbf{System schematic and forward predictions for the 
    cancer biomarker example.} 
    (a) Prostate cancer stem cells ($S$) divide at rate $\lambda$, producing either two stem cells or a stem and a differentiated cell ($D$), with probabilities governed by $p$. Differentiated cells ($D$) produce PSA at rate $\rho$, while the PSA concentration ($P$) decays at rate $\varphi$. When treatment is on, differentiated cells additionally decay at rate $\alpha$.
    (b) Median predictions and 95\% credible intervals under the no-pooling (blue) and partial-pooling (red) models for selected patients. Observations (filled circles) before the vertical dotted line are used for parameter inference, while PSA dynamics after the line are forward predictions. Left: all but the final treatment cycles are used for inference. Right: only the first treatment cycle is used for inference.
    }
    \label{fig:prostate-cancer-example}
  \end{figure}

Prostate cancer is one of the leading causes of cancer-related mortality among men worldwide \cite{litwin2017diagnosis}.
A commonly used treatment for advanced prostate cancer is androgen deprivation therapy (ADT).
However, a possible complication of continuous ADT is the emergence of treatment-resistant disease, which can lead to cancer progression and metastasis \cite{feldman2001development}.
One approach to impede the emergence of resistance is intermittent ADT, in which treatment is paused when a biomarker falls below a predefined threshold and restarted when it rises above the threshold again.  
In mathematical models of tumour dynamics, this treatment strategy is often interpreted through competition between drug-sensitive and resistant cancer cell populations. Under this framework, treatment-free periods allow sensitive populations to recover, which may slow the expansion of resistant disease.
We consider such a model presented by \textcite{brady2020prostate} using data from a previous clinical study \cite{Bruchovsky2006}. 
The model describes two competing cancer cell populations: resistant prostate cancer stem cells ($S$) and androgen-dependent differentiated cancer cells ($D$), together with the serum concentration of prostate-specific antigen (PSA), denoted by $P$, which is used as a biomarker to monitor treatment response and guide when treatment is paused or resumed.
The system dynamics are illustrated in Fig.~\ref{fig:prostate-cancer-example}a and are described by 
\begin{linenomath}
\begin{subequations}\label{eq:prostate-cancer-model}
\begin{align}
    \frac{dS(t)}{dt} &= \left( \frac{S(t)}{S(t)+D(t)} \right) p \lambda S(t), \\
    \frac{dD(t)}{dt} &= \left( 1- p\frac{S(t)}{S(t)+D(t)}  \right) \lambda S(t) - \alpha D(t) T, \\
    \frac{dP(t)}{dt} &= \rho D(t) - \varphi P(t),
\end{align}
\end{subequations}
\end{linenomath}
with initial conditions
\begin{equation}
S(0) = S_0, \qquad D(0) = D_0, \qquad P(0) = P_0.
\end{equation}
Here, $\lambda$ denotes the division rate of S-cells. With probability $p$, an S-cell divides symmetrically to produce two S-cells, whereas with probability $(1-p)$ it divides asymmetrically to produce one S-cell and one D-cell. 
The factor $\frac{S(t)}{S(t)+D(t)}$ represents a modelled negative feedback, whereby a larger relative abundance of D-cells corresponds to reduced S-cell proliferation.
The parameters $\rho$ and $\varphi$ denote the production and decay rates of PSA, respectively. Treatment status is represented by the binary indicator $T$, which specifies whether the patient is receiving ADT. When treatment is active, $T=1$ and D-cells decay at a rate $\alpha$.

For this model, we consider both no pooling and partial pooling to analyse patient biomarker dynamics.
The PSA measurements $\tilde{y}_{j,i}$ for patient $j$ at observation times $t_i$ are assumed to follow a normal distribution with both additive $(\sigma_j)$ and proportional $(\sigma_j')$ noise terms such that
\begin{equation}
    \tilde y_{j,i} = y(t_i;\xi_j) + \epsilon_{j,i}, \qquad 
    \epsilon_{j,i} \sim \mathcal{N}\left(0,\left(\sigma_j + y(t_i;\xi_j)\sigma'_j\right)^2\right),
\end{equation}
where $y(t_i;\xi_j)$ denotes the ODE output at time $t_i$ given parameters $\xi_j$. 
With fixed initial conditions, the parameters to infer are thus $\theta_j  = \left( p_{j},\lambda_j,\alpha_j,\rho_j,\varphi_j, \sigma_j, \sigma'_j \right)$. 
In the no-pooling case, these parameters are inferred independently for each patient, whereas in the partial-pooling case they are assumed to arise from shared population-level distributions.

To generate forward predictions of PSA dynamics, we consider ten patients and two inference scenarios: (i) when parameters are inferred using data from all but the final treatment cycle, and (ii) when parameters are inferred using data from only the first ADT treatment cycle. 
In case (i), both pooling strategies produce similar predictions because sufficient patient-specific data are available. However, in case (ii), when only observations from the first treatment cycle are available for parameter inference, the difference between the methods becomes more pronounced, with partial pooling producing more accurate predictions due to information sharing across patients.
This effect is visible in Fig.~\ref{fig:prostate-cancer-example}, where the 95\% predictive credible interval is substantially wider under no pooling for case (ii), but not case (i).
This comparison is further supported by Table~\ref{tab:biomarker-elpd}, which reports the $\widehat{\mathrm{elpd}}_{\mathrm{LOO}}$ values for both cases. 
When model parameters are inferred on all but the last treatment cycle, the total $\widehat{\mathrm{elpd}}_{\mathrm{LOO}}$ values are nearly identical between pooling strategies. 
In contrast, when model parameters are inferred only the first treatment cycle, the total $\widehat{\mathrm{elpd}}_{\mathrm{LOO}}$ favours partial pooling (\texttt{elpd\_diff} = -9.9, \texttt{se\_diff} = 3.6).  
While the individual $\widehat{\mathrm{elpd}}_{\mathrm{LOO}}$ values differ only slightly between pooling strategies for most patients, their aggregate difference supports the advantage of partial pooling in this setting. 
Prediction plots for all patients, together with diagnostic plots for $\widehat{\mathrm{elpd}}_{\mathrm{LOO}}$, are provided in the Supplementary Material (SM4), which also shows the corresponding results for complete pooling.

\begin{table}[h!]
     \caption{Comparison of $\widehat{\mathrm{elpd}}_{\mathrm{LOO}}$ values for patients (columns) under no pooling (No) and partial pooling (Part.). Higher values indicate better predictive performance, with the highest score for each patient shown in bold. }
    \label{tab:biomarker-elpd}
    \centering
    \resizebox{\textwidth}{!}{
    \begin{small}
            \begin{tabular}{lcccccccccc||c}
            \toprule
             & \textbf{P1} & \textbf{P4} & \textbf{P13} & \textbf{P14} & \textbf{P15} & \textbf{P17} & \textbf{P24} & \textbf{P26} & \textbf{P28} & \textbf{P29} & \textbf{Total} \\
            \hline
            \rowcolor{lightgray!40} \multicolumn{12}{c}{\textbf{(i) Parameter inference based on all but the last treatment cycle}} \\
            \textbf{No} & \textbf{-105.88} & -54.97 & \textbf{-91.60} & -58.90 & \textbf{-74.26} & -44.42 & \textbf{-56.73} & -59.61 & -141.58 & \textbf{-62.75}  & -750.7 \\
            \textbf{Part.} & -107.29 & \textbf{-52.80} & -92.76 & \textbf{-56.38} & -74.48 & \textbf{-43.76} & -57.66 & \textbf{-58.01} & \textbf{-141.48} & -66.08 & -750.7 \\
            \hline
            \rowcolor{lightgray!40} \multicolumn{12}{c}{\textbf{(ii) Parameter inference based on the first treatment cycle only}} \\
            \textbf{No}      & \textbf{-17.01} & -10.57 & -10.44 & -17.00 & -25.09              & -23.10 & -3.17                              & -9.77 & -46.49 & \textbf{-10.62 } & -173.3  \\
            \textbf{Part.} &  -17.78 & \textbf{-9.47} & \textbf{-7.85} & \textbf{-15.41} & \textbf{-24.24} & \textbf{-21.56} & \textbf{-1.96} & \textbf{-8.19} & \textbf{-44.95} & -11.99 & \textbf{-163.4}  \\
            \hline
            \bottomrule
            \end{tabular}
    \end{small}
    }
\end{table}

\newpage

\section{Conclusion}
This article provides a self-contained walkthrough of how to implement ODE models in Stan. 
Its discussion is focused, but not limited to, dynamical systems in biology.
Starting from Bayes’ rule and continuing with Monte Carlo–based sampling methods and, finally, computational inference in Stan, we have built up from theory to fully implemented ODE models with real-world data. 
Although we focused on first-order ODE systems, the same principles extend to more complex dynamical systems, including higher-order ODEs and partial differential equations.  
Such extensions typically require additional code development, or coupling Stan to external numerical solvers, and are therefore outside the scope of this article. 
A related limitation is that Stan natively requires an explicit likelihood function, which makes it less directly applicable to likelihood-free inference settings (e.g., agent-based models), although such cases can often be addressed using approximate likelihood approaches such as synthetic likelihoods or surrogate models.
For ODE models, however, this article is intended to lower the entry barrier to Bayesian analysis of dynamical systems and to demonstrate how Stan makes advanced computational inference methods accessible to modellers. 
This is particularly useful when modelling biological systems, since uncertainty is inherent to biology. 
The Bayesian perspective allows us to embrace and quantify this uncertainty in a principled and informative way.

\vspace{-.1cm}
\subsection*{Code access}
\vspace{-.1cm}
\noindent All code and data files are available in their original formats on the code-hosting platform GitHub: \url{https://github.com/jodie-c/stan_tutorial}.

\vspace{-.1cm}
\subsection*{Funding}
\vspace{-.1cm}
This work was partially supported by the Wallenberg AI, Autonomous Systems and
Software Program (WASP) funded by the Knut and Alice Wallenberg Foundation, which supported SH and CCG. 
The work was also partially supported by the Kjell och M{\"a}rta Beijer Foundation, which supported SH and JAC. 
In addition, SH was supported by Wenner-Gren Stiftelserna/the Wenner-Gren Foundations (WGF2022-0044) and the Swedish Research Council (project 2024-05621).

\vspace{-.1cm}
\subsection*{Author contributions}
\vspace{-.1cm}
SH, JAC: Conceptualisation, writing, implementation, data analysis, visualisation and supervision.

\noindent JF, CCG: Implementation and data analysis for Section 4.
\vspace{-.1cm}
\subsection*{Statements and Declarations:} 
\vspace{-.1cm}
The authors have no competing interests to declare that are relevant to the content of this article.

\vspace{-.2cm}
\enlargethispage{1\baselineskip}
\let\oldthebibliography\thebibliography
\let\endoldthebibliography\endthebibliography
\renewenvironment{thebibliography}[1]{
  \begin{oldthebibliography}{#1}
    \setlength{\itemsep}{0em}
    \setlength{\parskip}{0em}
}
{
  \end{oldthebibliography}
}
\enlargethispage{1\baselineskip}
\begin{spacing}{0.85}
\renewbibmacro{in:}{,}
\section*{References}
\vspace{-.1cm}
\AtNextBibliography{\small}
\printbibliography[heading=none]

@article{Pepin2025,
   author = {Kim M. Pepin and Keith Carlisle and Richard B. Chipman and Dana Cole and Dean P. Anderson and Michael G. Baker and Jackie Benschop and Michael Bunce and Rachelle N. Binny and Nigel French and Suzie Greenhalgh and Dion R.J. O’Neale and Scott McDougall and Fraser J. Morgan and Petra Muellner and Emil Murphy and Michael J. Plank and Daniel M. Tompkins and David T.S. Hayman},
   doi = {10.1057/s41599-025-05077-3},
   issue = {1},
   journal = {Humanities and Social Sciences Communications},
   title = {Practitioner perspectives on informing decisions in One Health sectors with predictive models},
   volume = {12},
   year = {2025}
}

@article{pugh_bib_2025,
	title = {A bibliometric study on mathematical oncology: interdisciplinarity, internationality, collaboration and trending topics},
	volume = {87},
	doi = {10.1007/s11538-025-01544-9},
	number = {12},
	journal = {Bulletin of Mathematical Biology},
	author = {Kira Pugh and Linnea Gyllingberg and Stanislav Stratiev and Sara Hamis},
	month = dec,
	year = {2025},
	pages = {174},
}

@article{Hoegh2020,
    author = {Andrew Hoegh},
    title = {Why Bayesian Ideas Should Be Introduced in the Statistics Curricula and How to Do So},
    journal = {Journal of Statistics Education},
    volume = {28},
    number = {3},
    pages = {222-228},
    year = {2020},
    publisher = {Taylor & Francis},
    doi = {10.1080/10691898.2020.1841591}
}

@article{Ferriera2020,
    title = {Theory and practical use of Bayesian methods in interpreting clinical trial data: a narrative review},
    journal = {British Journal of Anaesthesia},
    volume = {125},
    number = {2},
    pages = {201-207},
    year = {2020},
    doi = {10.1016/j.bja.2020.04.092},
    author = {David Ferreira and Mael Barthoulot and Julien Pottecher and Klaus D. Torp and Pierre Diemunsch and Nicolas Meyer}
}

@article{Muehlemann2023,
	Author = {Muehlemann, Natalia and Zhou, Tianjian and Mukherjee, Rajat and Hossain, Munshi Imran and Roychoudhury, Satrajit and Russek-Cohen, Estelle},
	Doi = {10.1007/s43441-023-00515-3},
	Journal = {Therapeutic Innovation \& Regulatory Science},
	Number = {3},
	Pages = {402--416},
	Title = {A Tutorial on Modern Bayesian Methods in Clinical Trials},
	Volume = {57},
	Year = {2023}
}

@article{Clark2023,
	Author = {Clark, Jennifer and Muhlemann, Natalia and Natanegara, Fanni and Hartley, Andrew and Wenkert, Deborah and Wang, Fei and Harrell, Frank E. and Bray, Ross},
	Doi = {10.1007/s43441-021-00357-x},
	Journal = {Therapeutic Innovation \& Regulatory Science},
	Number = {3},
	Pages = {417--425},
	Title = {Why are not There More Bayesian Clinical Trials? Perceived Barriers and Educational Preferences Among Medical Researchers Involved in Drug Development},
	Volume = {57},
	Year = {2023}
}

@article{Margossian2022,
   author = {Charles C. Margossian and Yi Zhang and William R. Gillespie},
   doi = {10.1002/psp4.12812},
   issue = {9},
   journal = {CPT: Pharmacometrics and Systems Pharmacology},
   title = {Flexible and efficient Bayesian pharmacometrics modeling using Stan and Torsten, Part I},
   volume = {11},
   year = {2022}
}

@article{Grinsztajn2021,
   author = {Léo Grinsztajn and Elizaveta Semenova and Charles C. Margossian and Julien Riou},
   doi = {10.1002/sim.9164},
   issue = {27},
   journal = {Statistics in Medicine},
   title = {Bayesian workflow for disease transmission modeling in Stan},
   volume = {40},
   year = {2021}
}

@article{Chatzilena2019,
   author = {Anastasia Chatzilena and Edwin van Leeuwen and Oliver Ratmann and Marc Baguelin and Nikolaos Demiris},
   doi = {10.1016/j.epidem.2019.100367},
   journal = {Epidemics},
   title = {Contemporary statistical inference for infectious disease models using Stan},
   volume = {29},
   year = {2019}
}

@article{MetropolisUlam1949,
  author  = {Metropolis, Nicholas and Ulam, S.},
  title   = {The Monte Carlo Method},
  journal = {Journal of the American Statistical Association},
  volume  = {44},
  number  = {247},
  pages   = {335--341},
  year    = {1949},
  month   = sep,
  doi     = {10.2307/2280232}
}

@article{Tierney1986,
   author = {Luke Tierney and Joseph B. Kadane},
   doi = {10.1080/01621459.1986.10478240},
   issn = {1537274X},
   issue = {393},
   journal = {Journal of the American Statistical Association},
   title = {Accurate approximations for posterior moments and marginal densities},
   volume = {81},
   year = {1986}
}

@article{jordan1999introduction,
   author = {Michael I. Jordan and Zoubin Ghahramani and Tommi S. Jaakkola and Lawrence K. Saul},
   doi = {10.1023/A:1007665907178},
   issue = {2},
   journal = {Machine Learning},
   title = {Introduction to variational methods for graphical models},
   volume = {37},
   year = {1999}
}

@article{metropolis1953equation,
   author = {Nicholas Metropolis and Arianna W. Rosenbluth and Marshall N. Rosenbluth and Augusta H. Teller and Edward Teller},
   doi = {10.1063/1.1699114},
   issue = {6},
   journal = {The Journal of Chemical Physics},
   title = {Equation of state calculations by fast computing machines},
   volume = {21},
   year = {1953}
}

@article{hastings1970monte,
   author = {W. K. Hastings},
   doi = {10.1093/biomet/57.1.97},
   issue = {1},
   journal = {Biometrika},
   title = {Monte Carlo sampling methods using Markov chains and their applications},
   volume = {57},
   year = {1970}
}

@article{duane1987hybrid,
   author = {Simon Duane and A. D. Kennedy and Brian J. Pendleton and Duncan Roweth},
   doi = {10.1016/0370-2693(87)91197-X},
   issue = {2},
   journal = {Physics Letters B},
   title = {Hybrid Monte Carlo},
   volume = {195},
   year = {1987}
}

@article{hoffman2014no,
  title={The No-U-Turn sampler: adaptively setting path lengths in Hamiltonian Monte Carlo},
  author={Hoffman, Matthew D and Gelman, Andrew and others},
  journal={J. Mach. Learn. Res.},
  volume={15},
  number={1},
  pages={1593--1623},
  year={2014}
}

@article{Roberts2001,
   author = {Gareth O. Roberts and Jeffrey S. Rosenthal},
   doi = {10.1214/ss/1015346320},
   issue = {4},
   journal = {Statistical Science},
   title = {Optimal Scaling for Various Metropolis-Hastings Algorithms},
   volume = {16},
   year = {2001}
}

@article{Roberts1997,
   author = {G. O. Roberts and A. Gelman and W. R. Gilks},
   doi = {10.1214/aoap/1034625254},
   issue = {1},
   journal = {Annals of Applied Probability},
   title = {Weak convergence and optimal scaling of random walk Metropolis algorithms},
   volume = {7},
   year = {1997}
}

@article{gelman1992inference,
  title={Inference from iterative simulation using multiple sequences},
  author={Gelman, Andrew and Rubin, Donald B},
  journal={Statistical science},
  volume={7},
  number={4},
  pages={457--472},
  year={1992},
  doi = {10.1214/ss/1177011136},
}

@article{vehtari2017practical,
   author = {Aki Vehtari and Andrew Gelman and Jonah Gabry},
   doi = {10.1007/s11222-016-9696-4},
   issue = {5},
   journal = {Statistics and Computing},
   title = {Practical Bayesian model evaluation using leave-one-out cross-validation and WAIC},
   volume = {27},
   year = {2017},
   pages={1413--1432}
}

@article{Bruchovsky2006,
   author = {Nicholas Bruchovsky and Laurence Klotz and Juanita Crook and Shawn Malone and Charles Ludgate and W. James Morris and Martin E. Gleave and S. Larry Goldenberg},
   doi = {10.1002/cncr.21989},
   issue = {2},
   journal = {Cancer},
   title = {Final results of the Canadian prospective Phase II trial of intermittent androgen suppression for men in biochemical recurrence after radiotherapy for locally advanced prostate cancer: Clinical parameters},
   volume = {107},
   year = {2006}
}

@article{brown2025coral,
   author = {Kaitlyn Brown and Robyn P. Araujo and Paul Corry and Adrianne L. Jenner},
   doi = {10.1016/j.ecolmodel.2025.111110},
   journal = {Ecological Modelling},
   title = {Coral bleaching dynamics on the Great Barrier Reef: New insights from a mathematical model},
   volume = {505},
   year = {2025}
}

@article{hughes2003climate,
  title={Climate Change, Human Impacts, and the Resilience of Coral Reefs},
  author={Hughes, Terry P and Baird, Andrew H and Bellwood, David R and Card, Margaret and Connolly, Sean R and Folke, Carl and Grosberg, Richard and Hoegh-Guldberg, Ove and Jackson, Jeremy BC and Kleypas, Janice and others},
  journal={Science},
  volume={301},
  number={5635},
  pages={929--933},
  year={2003},
  publisher={American Association for the Advancement of Science},
  doi = {10.1126/science.1085046}
}

@article{feldman2001development,
  title={The development of androgen-independent prostate cancer},
  author={Feldman, Brian J and Feldman, David},
  journal={Nature Reviews Cancer},
  volume={1},
  number={1},
  pages={34--45},
  year={2001},
  publisher={Nature Publishing Group UK London},
  doi = {10.1038/35094009},
}

@article{vehtari2021rank,
  title={Rank-normalization, folding, and localization: An improved $\hat{R}$ for assessing convergence of MCMC (with discussion)},
  author={Vehtari, Aki and Gelman, Andrew and Simpson, Daniel and Carpenter, Bob and B{\"u}rkner, Paul-Christian},
  journal={Bayesian analysis},
  volume={16},
  number={2},
  pages={667--718},
  year={2021},
  publisher={International Society for Bayesian Analysis},
  doi = {10.1214/20-BA1221}
}

@article{Gelman2015,
   author = {Andrew Gelman and Daniel Lee and Jiqiang Guo},
   doi = {10.3102/1076998615606113},
   issn = {19351054},
   issue = {5},
   journal = {Journal of Educational and Behavioral Statistics},
   title = {Stan: A Probabilistic Programming Language for Bayesian Inference and Optimization},
   volume = {40},
   year = {2015}
}

@article{Carpenter2017,
   author = {Bob Carpenter and Andrew Gelman and Matthew D. Hoffman and Daniel Lee and Ben Goodrich and Michael Betancourt and Marcus A. Brubaker and Jiqiang Guo and Peter Li and Allen Riddell},
   doi = {10.18637/jss.v076.i01},
   issue = {1},
   journal = {Journal of Statistical Software},
   title = {Stan: A probabilistic programming language},
   volume = {76},
   year = {2017}
}

@article{Burkner2017,
   author = {Paul Christian Bürkner},
   doi = {10.18637/jss.v080.i01},
   journal = {Journal of Statistical Software},
   title = {brms: An R package for Bayesian multilevel models using Stan},
   volume = {80},
   year = {2017}
}

@article{litwin2017diagnosis,
  title={The diagnosis and treatment of prostate cancer: a review},
  author={Litwin, Mark S and Tan, Hung-Jui},
  journal={JAMA - Journal of the American Medical Association},
  volume={317},
  number={24},
  pages={2532--2542},
  year={2017},
  publisher={American Medical Association},
   doi = {10.1001/jama.2017.7248}
}

@article{brady2020prostate,
   author = {Renee Brady-Nicholls and John D. Nagy and Travis A. Gerke and Tian Zhang and Andrew Z. Wang and Jingsong Zhang and Robert A. Gatenby and Heiko Enderling},
   doi = {10.1038/s41467-020-15424-4},
   issue = {1},
   journal = {Nature Communications},
   title = {Prostate-specific antigen dynamics predict individual responses to intermittent androgen deprivation},
   volume = {11},
   year = {2020}
}

@misc{betancourt2017conceptual,
  author       = {Betancourt, Michael},
  title        = {A Conceptual Introduction to Hamiltonian Monte Carlo},
  year         = {2017},
  eprint       = {1701.02434},
  archivePrefix= {arXiv},
  primaryClass = {stat.ME}
}

@misc{murphy2007conjugate,
  title={Conjugate Bayesian analysis of the Gaussian distribution},
  author={Murphy, Kevin P},
  year={2007}
}

@inproceedings{Minka2001,
author = {Minka, Thomas P.},
title = {Expectation propagation for approximate Bayesian inference},
year = {2001},
isbn = {1558608001},
publisher = {Morgan Kaufmann Publishers Inc.},
address = {San Francisco, CA, USA},
booktitle = {Proceedings of the Seventeenth Conference on Uncertainty in Artificial Intelligence},
pages = {362–369},
numpages = {8},
location = {Seattle, Washington},
series = {UAI'01}
}

@book{Lambert2018,
    title = {A Student’s Guide to Bayesian Statistics (1st ed.)},
    author = {Lambert, Ben},
    isbn = {9781473916364},
    year = {2018},  
    publisher = {SAGE Publications},
    doi={10.4135/9781036234546}
}

@book{Fraenzi2015,
   author = {Fränzi Korner-Nievergelt and Tobias Roth and Stefanie von Felten and Jérôme Guélat and Bettina Almasi and Pius Korner-Nievergelt},
   doi = {10.1016/C2013-0-23227-X},
   isbn = {9780128016787},
   journal = {Bayesian Data Analysis in Ecology Using Linear Models with R, BUGS, and Stan},
   month = {4},
   pages = {1-316},
   publisher = {Elsevier Inc.},
   title = {Bayesian Data Analysis in Ecology Using Linear Models with R, BUGS, and Stan},
   year = {2015}
}

@inbook{neal2011mcmc,
   author = {Radford M. Neal},
   doi = {10.1201/b10905-6},
   booktitle = {Handbook of Markov Chain Monte Carlo},
   title = {MCMC using hamiltonian dynamics},
   year = {2011},
   editor    = {Steve Brooks and Andrew Gelman and Galin L. Jones and Xiao-Li Meng},
   publisher = {Chapman and Hall/CRC},
}

@book{RobertCasella2004,
  author    = {Robert, Christian P. and Casella, George},
  title     = {Monte Carlo Statistical Methods},
  year      = {2004},
  publisher = {Springer-Verlag New York},
  address   = {New York},
  edition   = {2}
}

@article{betancourt2015hamiltonian,
  title={Hamiltonian Monte Carlo for Hierarchical Models},
  author={Betancourt, Michael and Girolami, Mark},
  journal={Current Trends in Bayesian Methodology with Applications},
  pages={79},
  year={2015},
  publisher={CRC Press},
  doi = {10.1201/b18502-5}
}

@book{gelman1995bayesian,
   author = {Andrew Gelman and John B. Carlin and Hal S. Stern and David B. Dunson and Aki Vehtari and Donald B. Rubin},
   title = {Bayesian Data Analysis, Third edition},
   year = {2013},
   publisher={Chapman and Hall/CRC},
   doi= {10.1201/b16018}
}

@book{KampourakisMcCain2019,
    title = {How Uncertainty Makes Science Advance},
    author = {Kampourakis, Kostas and McCain, Kevin},
    isbn = {9780190871666},
    year = {2019},  
    publisher = {Oxford University Press},
    doi = {10.1007/978-1-4757-4145-2}
}

@inbook{McElreath2018,
   author = {Richard McElreath},
   doi = {10.1201/9781315372495-6},
   booktitle = {Statistical Rethinking: A Bayesian Course with Examples in R and Stan. },
   year = {2018}
}

@misc{stan,
    title = {Stan Modeling Language Users Guide and Reference Manual, VERSION 2.33},
    author = {{Stan Development Team}},
    year = {2022},
    url = {https://mc-stan.org/},
}

@misc{vehtari2015loo,
  title={loo: Efficient leave-one-out cross-validation and WAIC for Bayesian models},
  author={Vehtari, Aki and Gabry, Jonah and Magnusson, M{\aa}ns and Yao, Yuling and B{\"u}rkner, Paul-Christian and Paananen, Topi and Gelman, Andrew},
  year={2015},
  publisher={The R Foundation},
  url={https://cran.r-project.org/web/packages/loo/}
}

@misc{aims,
    title = {AIMS Long-term Monitoring Program: Video and Photo Transects (Great Barrier Reef)},
    author = {{Australian Institute of Marine Science (AIMS)}},
    year = {2015},
    note = {Accessed 13-Feb-2025}
}

@misc{MargossianStanCon2017,
  author       = {Margossian, Charles and Gillespie, Brandon},
  title        = {Differential Equation-Based Models in Stan},
  year         = {2017},
  doi          = {10.5281/zenodo.1284264},
  howpublished = {Zenodo}
}

@misc{stan2019,
    title = {Stan Modeling Language Users Guide and Reference Manual, VERSION 2.19.1},
    author = {{Stan Development Team}},
    year = {2019},
    url = {https://mc-stan.org/},
}

@misc{gelman2020bayesianworkflow,
      title={Bayesian Workflow}, 
      author={Andrew Gelman and Aki Vehtari and Daniel Simpson and Charles C. Margossian and Bob Carpenter and Yuling Yao and Lauren Kennedy and Jonah Gabry and Paul-Christian Bürkner and Martin Modrák},
      year={2020},
      eprint={2011.01808},
      archivePrefix={arXiv},
      primaryClass={stat.ME},
      url={https://arxiv.org/abs/2011.01808}, 
}

@article{Tierney1994,
   author = {Luke Tierney},
   journal = {Annals of Statistics},
   title = {Markov Chains for Exploring Posterior Distributions},
   volume = {22},
   year = {1994},
   doi = {10.1214/aos/1176325750}
}

\end{spacing}

\end{document}


\begin{Large}

\noindent{\it Supplementary Material to: }

\noindent A practical introduction to ODE modelling in Stan for biological systems
\end{Large}
\vspace{.5cm}

\noindent Sara Hamis$^1$, John Forslund$^1$, Cici Chen Gu$^1$, Jodie A. Cochrane$^1$.
\vspace{0.1cm}

\small{
\noindent $^1$Department of Information Technology, Uppsala University, Uppsala, Sweden.
}
\vspace{0.1cm}

\noindent Correspondence: sara.hamis@it.uu.se. 

\subsection*{SM1: Citation analysis of Stan in selected mathematical biology journals}
We identified articles in five well-established mathematical biology journals that cite Stan,
following the methods and journal selection of a recent bibliometric study \cite{pugh_bib_2025}. 
This includes both direct citations of the Stan programming framework and citations of other works that mention Stan in the title, up to the end of 2024.
The citing articles are listed in Table \ref{tab:stan_citations}. 
\begin{table}[h]
\centering
\begin{tabular}{l c l}
\hline
Citing article (DOI) & Year & Cited Stan-related reference \\
\hline
\multicolumn{3}{l}{\textbf{Bulletin of Mathematical Biology}} \\
\hline
\href{https://doi.org/10.1007/s11538-018-00556-y}{10.1007/s11538-018-00556-y} & 2019 & Gelman et al. (2015) \cite{Gelman2015} \\
\href{https://doi.org/10.1007/s11538-022-01088-2}{10.1007/s11538-022-01088-2} & 2022 & Carpenter et al. (2017) \cite{carpenter2017} \\
\href{https://doi.org/10.1007/s11538-023-01169-w}{10.1007/s11538-023-01169-w} & 2023 & McElreath (2018) \cite{McElreath2018} \\
\href{https://doi.org/10.1007/s11538-024-01285-1}{10.1007/s11538-023-01169-w} & 2024 & Margossian \& Gillespie, 2017 \cite{MargossianStanCon2017} ; Chatzilena et al. (2019) \cite{Chatzilena2019} \\
\href{https://doi.org/10.1007/s11538-024-01326-9}{10.1007/s11538-024-01326-9} & 2024 & Grinsztajn et al. (2021) \cite{Grinsztajn2021} \\
\hline
\multicolumn{3}{l}{\textbf{Journal of Mathematical Biology}} \\
\hline
\href{https://doi.org/10.1007/s00285-021-01669-0}{10.1007/s00285-021-01669-0} & 2021 & Carpenter et al. (2017) \cite{carpenter2017} \\
\href{https://doi.org/10.1007/s00285-022-01739-x}{10.1007/s00285-022-01739-x} & 2022 & Carpenter et al. (2017) \cite{carpenter2017} \& Stan Development team (2019) \cite{stan2019} \\
\hline
\multicolumn{3}{l}{\textbf{Journal of Theoretical Biology}} \\
\hline
\href{https://doi.org/10.1016/j.jtbi.2018.03.026}{10.1016/j.jtbi.2018.03.026} & 2018 & Carpenter et al. (2017) \cite{carpenter2017} \\
\href{https://doi.org/10.1016/j.jtbi.2018.04.006}{10.1016/j.jtbi.2018.04.006} & 2018 & Gelman et al. (2015) \cite{Gelman2015} \\
\href{https://doi.org/10.1016/j.jtbi.2024.111793}{10.1016/j.jtbi.2024.111793} & 2024 & Carpenter et al. (2017) \cite{carpenter2017} \\
\hline
\multicolumn{3}{l}{\textbf{Mathematical Biosciences}} \\
\hline
\href{https://doi.org/10.1016/j.mbs.2020.108438}{10.1016/j.mbs.2020.108438} & 2020 & McElreath (2018) \cite{McElreath2018} \\
\href{https://doi.org/10.1016/j.mbs.2021.108614}{10.1016/j.mbs.2021.108614} & 2021 & Chatzilena et al. (2019)\cite{Chatzilena2019} \\
\hline
\multicolumn{3}{l}{\textbf{Mathematical Biosciences and Engineering}} \\
\hline
\href{https://doi.org/10.3934/mbe.2022323}{10.3934/mbe.2022323} & 2022 & Carpenter et al. (2017) \citep{carpenter2017} \\
\href{https://doi.org/10.3934/mbe.2023207}{10.3934/mbe.2023207} & 2023 & Gelman et al. (2015) \cite{Gelman2015} \\
\href{https://doi.org/10.3934/mbe.2023488}{10.3934/mbe.2023488} & 2023 & B{\"u}rkner (2017) \cite{Burkner2017} \\
\hline
\end{tabular}
\caption{All Stan-related references cited in five mathematical biology journals up to the end of 2024.}
\label{tab:stan_citations}
\end{table}

\vspace{-.2cm}
\subsection*{SM2: Code for the toy model example}
This section contains the Stan code used for the toy model, 
as well as minimal R codes for interfacing with Stan and inspecting Stan outputs. 
For prettified result plots (as shown in the main article), we refer to the project's GitHub page,
which also contains codes for the presented real-world examples.

\subsection*{Stan code for the toy model}
\begin{lstlisting}[caption=The full Stan code for the toy model., label={lst:listing}, style=stan]
functions {
  vector logistic_ode(real t, vector y, vector p) {
    vector[2] dydt;
    real r1 = p[1];
    real r2 = p[2];
    real K  = p[3];
    
    // ---- ODE model ----
    dydt[1] = r1 * y[1] * (1 - (y[1] + y[2]) / K);
    dydt[2] = r2 * y[2] * (1 - (y[1] + y[2]) / K);

    return dydt;
  }
}

data {
  int<lower=1> no_ts;                           // Number of observation time points
  int<lower=1> no_wells;                        // Number of wells
  real t0;                                      // Initial time
  array[no_ts] real<lower=t0> ts;               // Observation times
  array[no_wells] matrix[2, no_ts] y_tilde;     // Observed data

  int<lower=1> no_ts_gen;                       // Number of prediction time points
  array[no_ts_gen] real ts_gen;                 // Prediction times
}

transformed data {
  //transformed data here
}

parameters {
  vector<lower=0>[3] p;                         // p[1]=r1, p[2]=r2, p[3]=K
  real<lower=0> sigma;                          // Observation noise
  vector<lower=0>[2] y0;                        // Initial values for both species
}

transformed parameters {
  vector[2] y0_state;                           // Initial state vector
  array[no_ts] vector[2] y_state;               // Deterministic ODE solution (solver output)
  matrix[2, no_ts] y;                           // Deterministic trajectories (for likelihood)

  // ---- Solve ODE system at observation times ----
  y0_state = y0;
  y_state = ode_rk45(logistic_ode, y0_state, t0, ts, p);

  // ---- Convert solver output to 2 x no_ts matrix ----
  for (i in 1:no_ts) {
    y[1, i] = y_state[i][1];
    y[2, i] = y_state[i][2];
  }
}

model {
  // ---- Priors ----
  p[1] ~ normal(0.2, 0.1);                      // Growth rate r1
  p[2] ~ normal(0.2, 0.1);                      // Growth rate r2
  p[3] ~ normal(5, 1);                          // Shared carrying capacity K
  sigma ~ normal(0, 0.5);                       // Noise standard deviation
  y0 ~ normal(1, 0.5);                          // Initial values

  // ---- Likelihood (elementwise) ----
  for (j in 1:no_wells) {
    to_vector(y_tilde[j]) ~ normal(to_vector(y), sigma); 
  }
}

generated quantities {
  // ---- Predictions at time points ts_gen ----
  array[no_ts_gen] vector[2] y_state_pred;      // Deterministic ODE solution (solver output)
  matrix[2, no_ts_gen] y_mean;                  // Noise-free trajectory 
  matrix[2, no_ts_gen] y_pred;                  // Posterior predictive draws (with noise)

  vector[2] y0_state_gen;
  y0_state_gen = y0;

  // ---- Solve ODE at prediction times ----
  y_state_pred = ode_rk45(logistic_ode, y0_state_gen, t0, ts_gen, p);

  // ---- Store noise-free mean + generate noisy observations ----
  for (i in 1:no_ts_gen) {
    y_mean[1, i] = y_state_pred[i][1];         // Noise-free ODE solution
    y_mean[2, i] = y_state_pred[i][2];                 

    y_pred[1, i] = y_mean[1, i] + normal_rng(0, sigma); // Noisy prediction
    y_pred[2, i] = y_mean[2, i] + normal_rng(0, sigma); 
  }
}
\end{lstlisting}

\subsection*{R code for the toy model presented in the main article.}

\begin{lstlisting}[caption=The R code for the toy model presented in the main article., label={lst:listing}, style=stan]
# ============================================================
# Load required packages and set random seed
# ============================================================
library(rstan)
library(posterior)
library(bayesplot)
library(ggplot2)
library(dplyr)
library(tidyr)
set.seed(123)

# ============================================================
# Generate fake data (2 competing species with Gaussian noise)
# ============================================================

# True parameter values (matching the Stan parameterization:
# p[1] = r1, p[2] = r2, p[3] = K)
r1_true    <- 0.15
r2_true    <- 0.10
K_true     <- 4.0
y0_true    <- c(1.2, 0.8)   # initial abundances for the two species
sigma_true <- 0.2           # observation noise standard deviation

# Experimental design: replicate wells and observation times
no_wells <- 6
t0 <- 0
ts <- seq(0, 48, by = 6)
if (ts[1] <= t0) ts[1] <- t0 + 1e-8
no_ts <- length(ts)

# Dense time grid used for posterior predictive visualization
ts_gen <- seq(min(ts), max(ts), length.out = 200)
no_ts_gen <- length(ts_gen)

# Deterministic (noise-free) trajectories for the two-species competition model.
# There is no simple closed-form solution when r1 != r2, so we simulate the
# coupled ODE system numerically using a fourth-order Runge-Kutta integrator.

compete_rhs <- function(t, y, r1, r2, K) {
  y1 <- y[1]
  y2 <- y[2]
  
  c(
    r1 * y1 * (1 - (y1 + y2) / K),
    r2 * y2 * (1 - (y1 + y2) / K)
  )
}

rk4_sim <- function(times, y0, r1, r2, K) {
  y <- matrix(NA_real_, nrow = length(times), ncol = 2)
  y[1, ] <- y0
  
  for (i in 2:length(times)) {
    dt <- times[i] - times[i - 1]
    t  <- times[i - 1]
    yi <- y[i - 1, ]
    
    k1 <- compete_rhs(t,          yi,             r1, r2, K)
    k2 <- compete_rhs(t + 0.5*dt, yi + 0.5*dt*k1, r1, r2, K)
    k3 <- compete_rhs(t + 0.5*dt, yi + 0.5*dt*k2, r1, r2, K)
    k4 <- compete_rhs(t + dt,     yi + dt*k3,     r1, r2, K)
    
    y[i, ] <- yi + (dt / 6) * (k1 + 2*k2 + 2*k3 + k4)
    y[i, ] <- pmax(y[i, ], 0)  # enforce nonnegative abundances
  }
  y
}

# Noise-free trajectories evaluated at the observation times (dimension: no_ts x 2)
mu_true_ts <- rk4_sim(ts, y0_true, r1_true, r2_true, K_true)

# Plot noise-free trajectories
plot(ts, mu_true_ts[, 1], type = "l", lwd = 2,
     xlab = "time", ylab = "population",
     ylim = c(0, 4.5), xlim = c(0, 48),
     main = "Simulated 2-species competition data")
lines(ts, mu_true_ts[, 2], lwd = 2)

# Generate replicate observations: each well is (mu + N(0, sigma)) for both species
# Stan expects: array[no_wells] matrix[2, no_ts]
y_tilde_list <- vector("list", no_wells)

for (w in 1:no_wells) {
  y1_obs <- rnorm(no_ts, mean = mu_true_ts[, 1], sd = sigma_true)
  y2_obs <- rnorm(no_ts, mean = mu_true_ts[, 2], sd = sigma_true)
  y_tilde_list[[w]] <- rbind(y1_obs, y2_obs)  # 2 x no_ts
  
  points(ts, y1_obs, pch = 16)
  points(ts, y2_obs, pch = 16)
}

# ============================================================
# Run Bayesian inference with Stan
# ============================================================

stan_file <- "~/toymodel.stan"

stan_data <- list(
  no_ts     = no_ts,
  no_wells  = no_wells,
  t0        = t0,
  ts        = ts,
  y_tilde   = y_tilde_list,
  no_ts_gen = no_ts_gen,
  ts_gen    = ts_gen
)

fit <- stan(
  file = stan_file,
  data = stan_data,
  chains = 4,
  iter = 2000,
  seed = 123
)

# ============================================================
# Extract draws from fit
# ============================================================
post <- rstan::extract(fit)

# ============================================================
# Print summary of fit
# ============================================================
print(fit, pars = c("p", "y0", "sigma"), probs = c(0.05, 0.5, 0.95))

# ============================================================
# Inspect pair plots
# ============================================================
mcmc_pairs(fit, pars = c("p[1]", "p[2]", "p[3]", "y0[1]", "y0[2]", "sigma"), 
           off_diag_args = list(size = 0.5, alpha = 0.2), 
           diag_fun = "dens", diag_args = list(alpha = 0.6))
\end{lstlisting}

\subsection*{SM3: Additional results for the coral reef example }

\begin{figure}[htp]
\includegraphics[width=\linewidth]{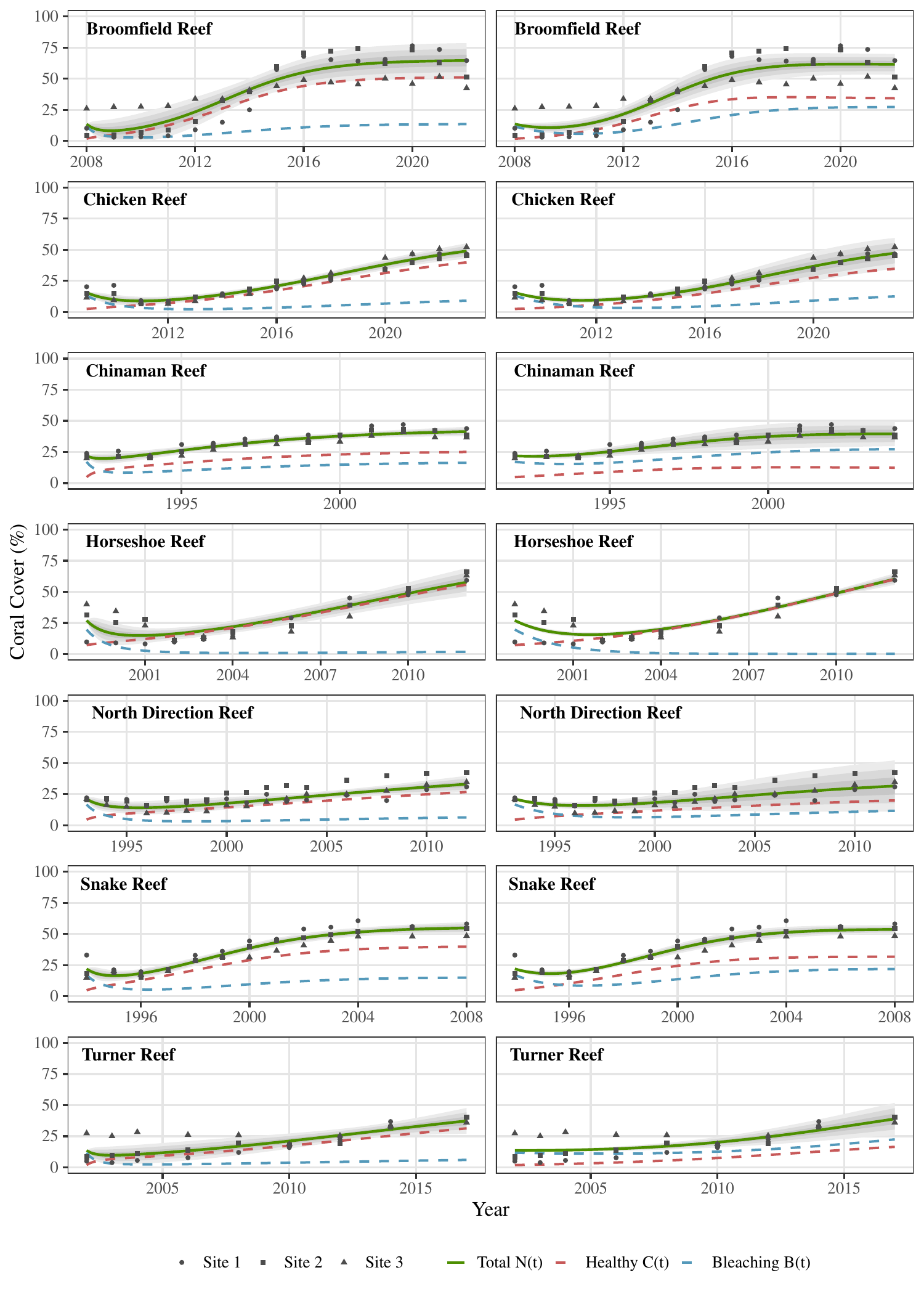}
    \caption{\textbf{Coral reef example cont.} Total coral reef coverage predictions for an additional seven reefs. 
    The left plots shows the results for the Bayesian analysis, and the right plots for the frequentist analysis. One, two, and three standard deviations are indicated by the grey shaded bands.}\label{fig:coral-reef-pred-supp}
  \end{figure}

\begin{figure}[htp]
\includegraphics[width=\linewidth]{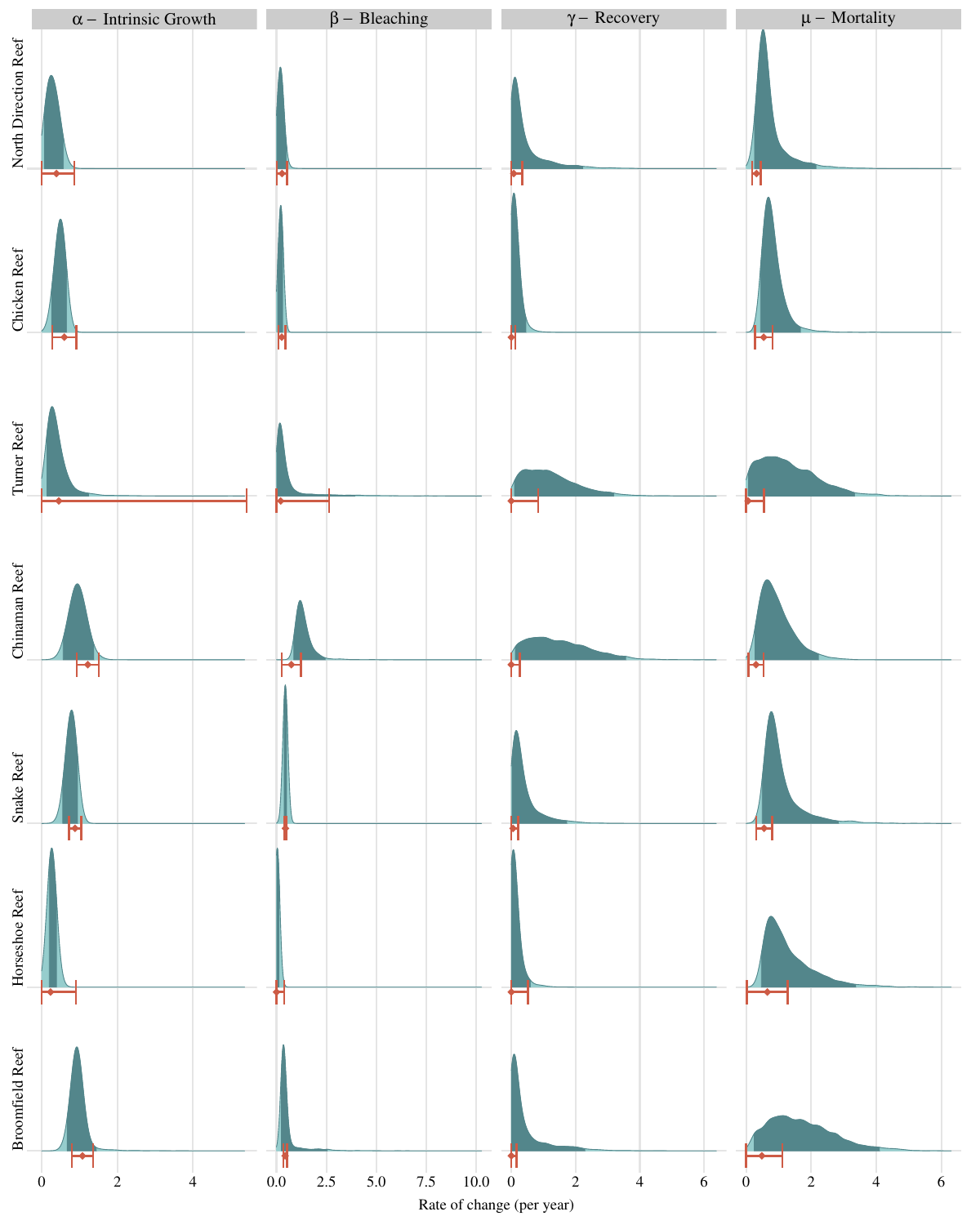}
    \caption{\textbf{Coral reef example cont.} Comparison between parameter estimates for frequentist and Bayesian methods for the seven additional reefs. For the frequentist approach, the 95\% confidence interval is shown, whereas the full posterior is shown for the Bayesian approach with 95\% credible interval shaded in dark blue.}\label{fig:coral-reef-param-supp}
  \end{figure}

Results for an additional 7 coral reefs are presented here. 
Predictions of total coral cover over time are shown in Fig.~\ref{fig:coral-reef-pred-supp}, while Fig.~\ref{fig:coral-reef-param-supp} shows the parameter estimate comparisons.

\newpage 
\subsection*{SM4: Additional results for the biomarker example }

\begin{figure}[htp]
 \vfill
    \begin{subfigure}[tb]{.49\linewidth}
      \centering
      \includegraphics[width=\linewidth]{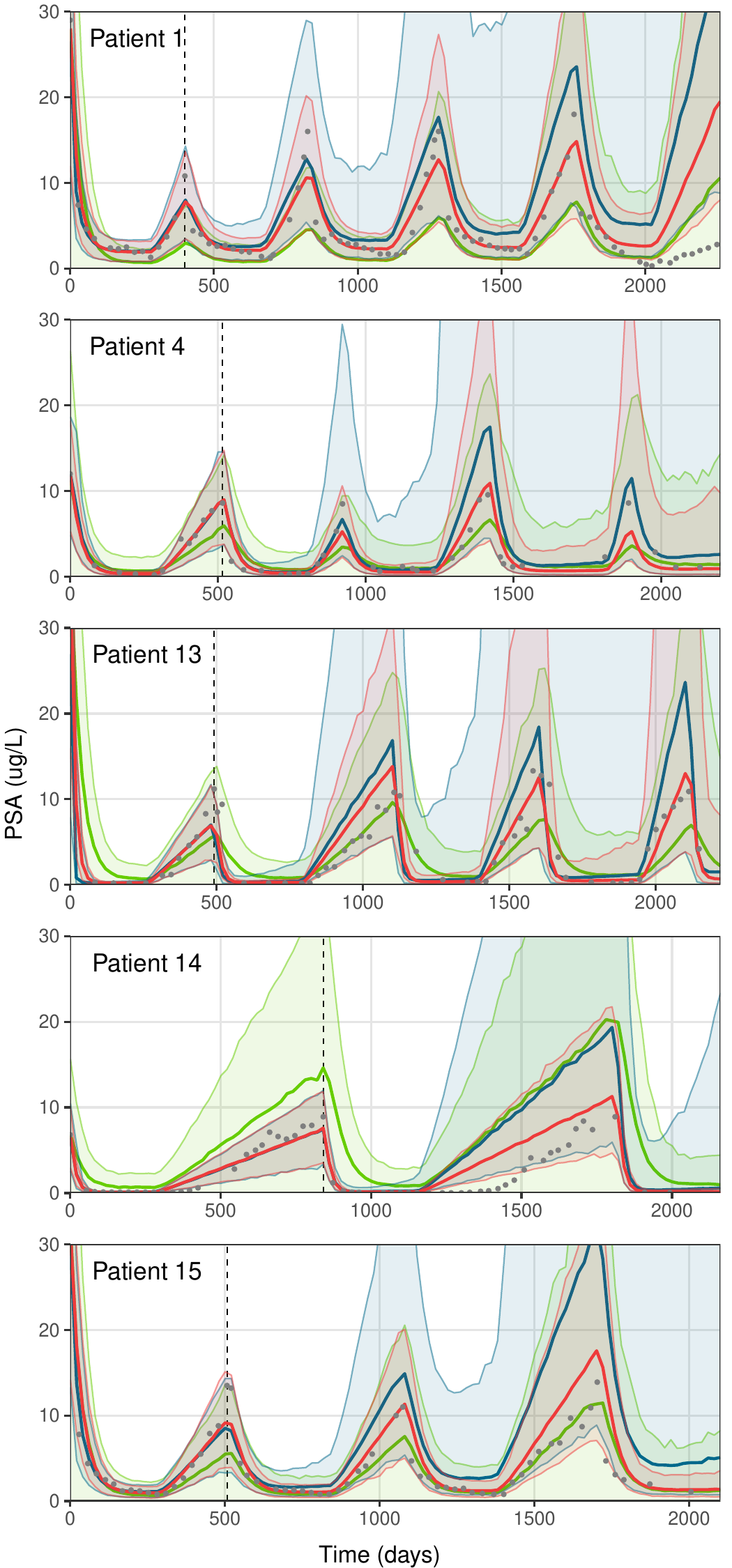}
    \end{subfigure}
    \hfill
    \begin{subfigure}[tb]{.49\linewidth}
      \centering
      \includegraphics[width=\linewidth]{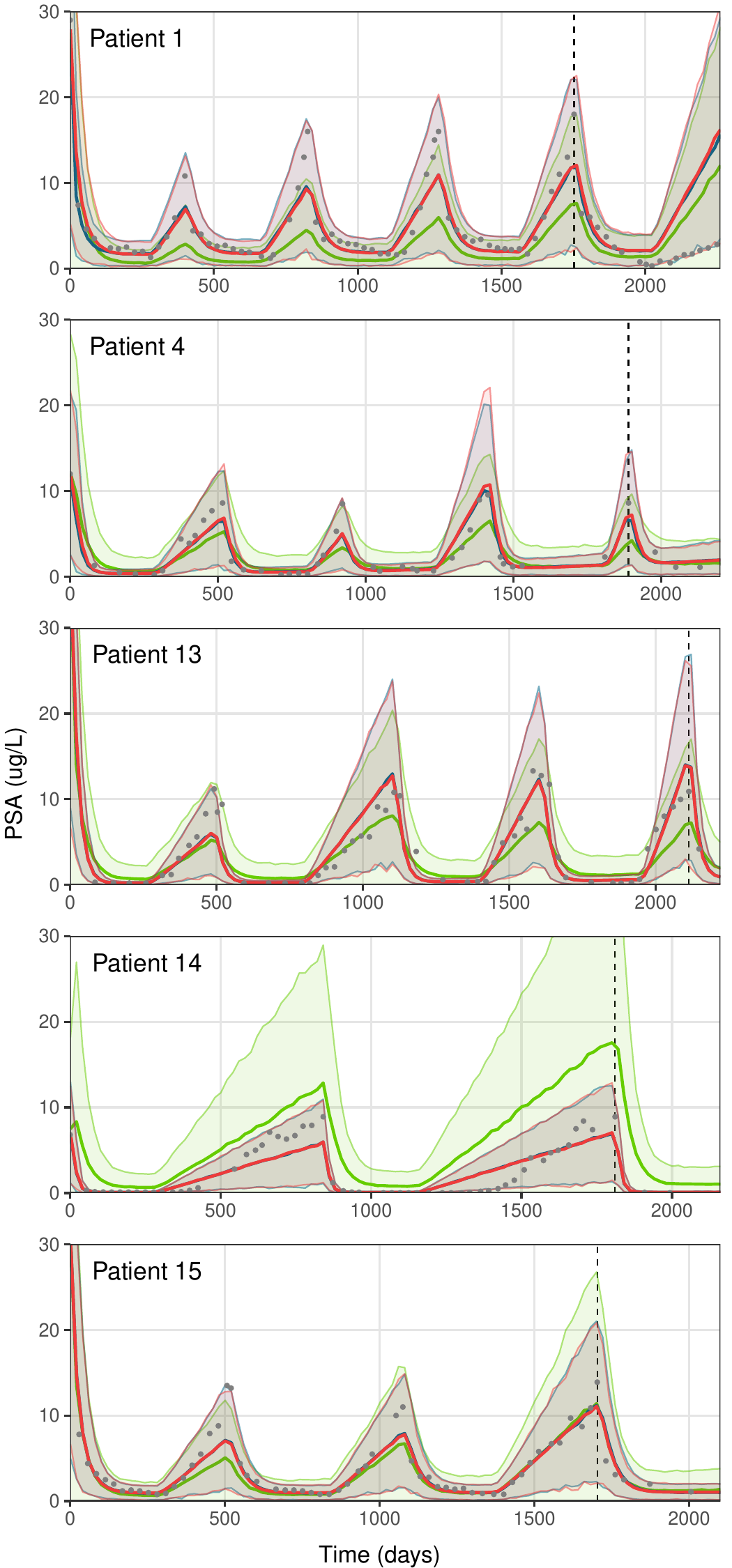}
    \end{subfigure}
    \caption{\textbf{Prostate cancer biomarker example cont.} Median and 95\% credible intervals for selected patients using complete pooling (green), no pooling (blue) and partial pooling (red) methods. Observations after the dotted line are unseen during inference.}\label{fig:biomarker-supp}
     \vfill
  \end{figure}

\begin{figure}[htp]
    \begin{subfigure}[tb]{.49\linewidth}
      \centering
      \includegraphics[width=\linewidth]{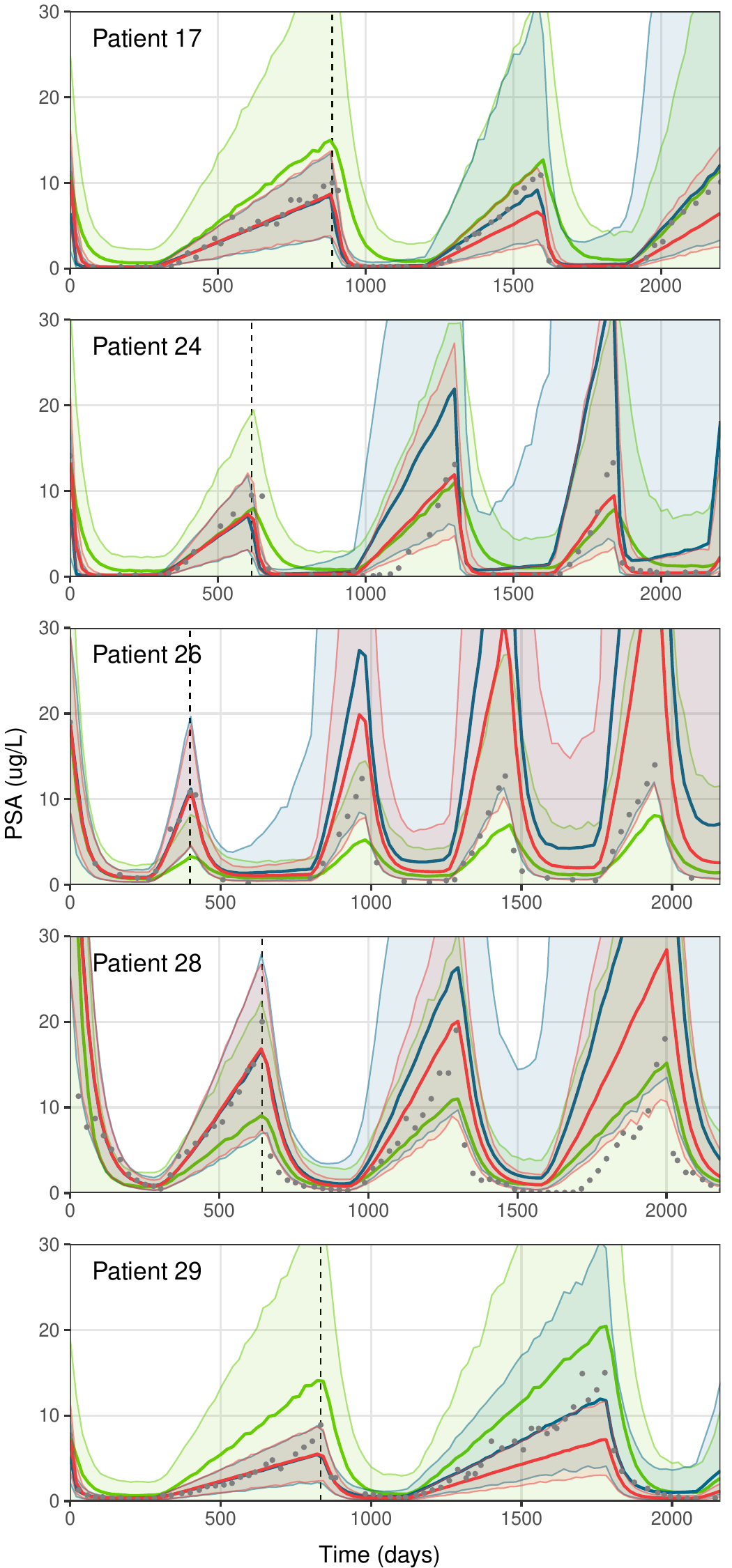}
    \end{subfigure}
    \hfill
    \begin{subfigure}[tb]{.49\linewidth}
      \centering
      \includegraphics[width=\linewidth]{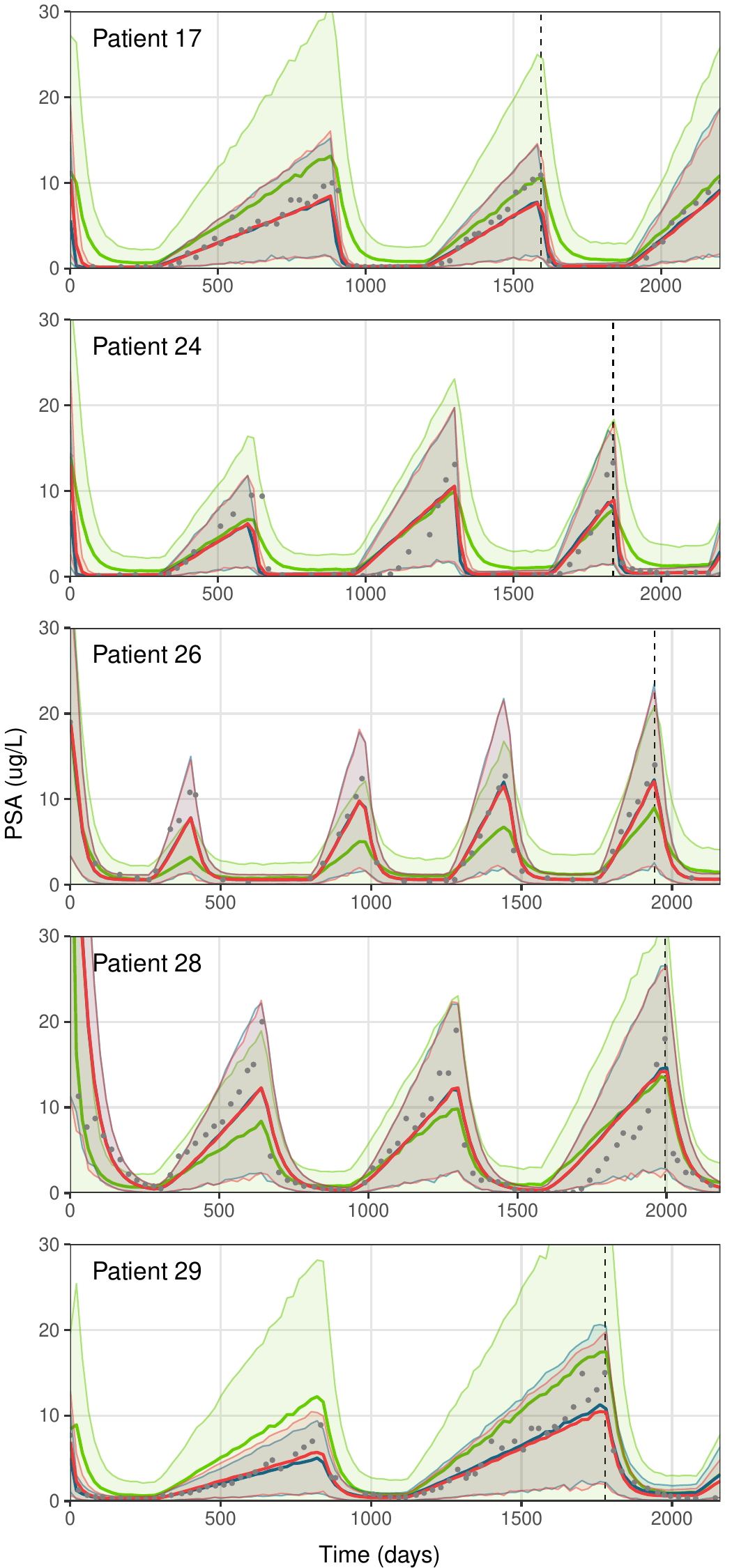}
    \end{subfigure}
    \caption{\textbf{Prostate cancer biomarker example cont.}  Median and 95\% credible intervals for selected patients using complete pooling (green), no pooling (blue) and partial pooling (red) methods. Observations after the dotted line are unseen during inference.}\label{fig:biomarker-supp-2}
  \end{figure}

Fig.~\ref{fig:biomarker-supp} and \ref{fig:biomarker-supp-2} show the posterior predictive distributions (including 95\% credible interval) for 10 patients when using complete, no, and partial pooling methods. Note that complete pooling resulted in a substantial number of divergences, but has been included here for completeness. 
\vspace{1.5cm}

\newpage

\begin{figure}[htp]
    \begin{subfigure}[tb]{.49\linewidth}
      \centering
      \includegraphics[width=\linewidth]{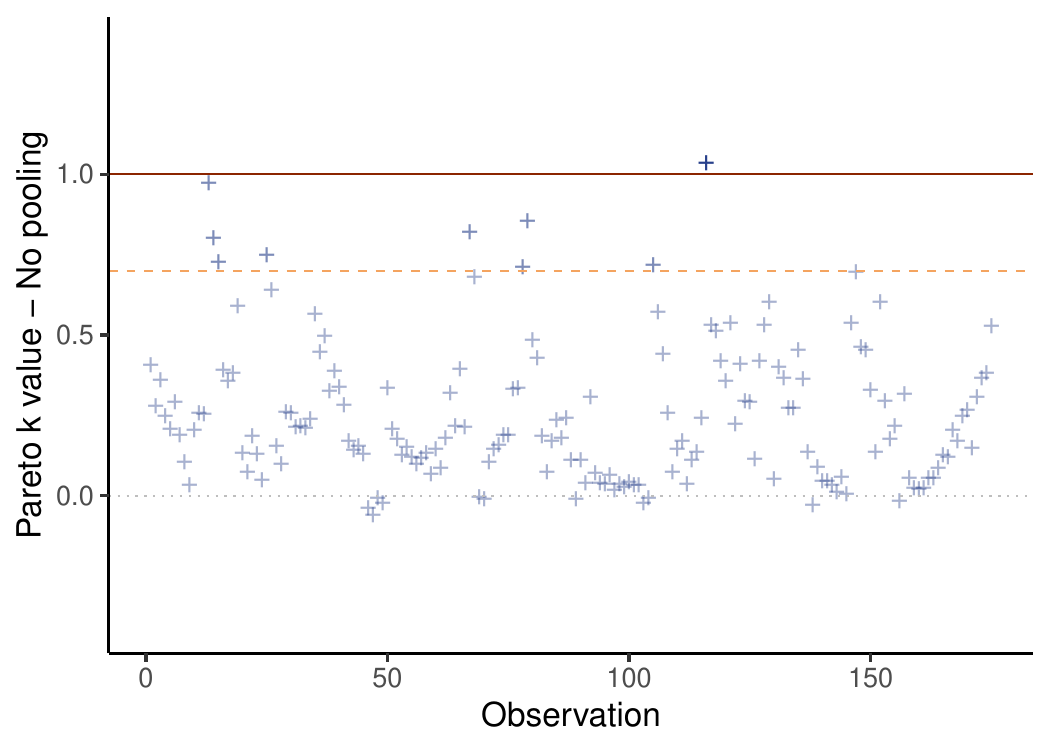}
    \end{subfigure}
    \hfill
    \begin{subfigure}[tb]{.49\linewidth}
      \centering
      \includegraphics[width=\linewidth]{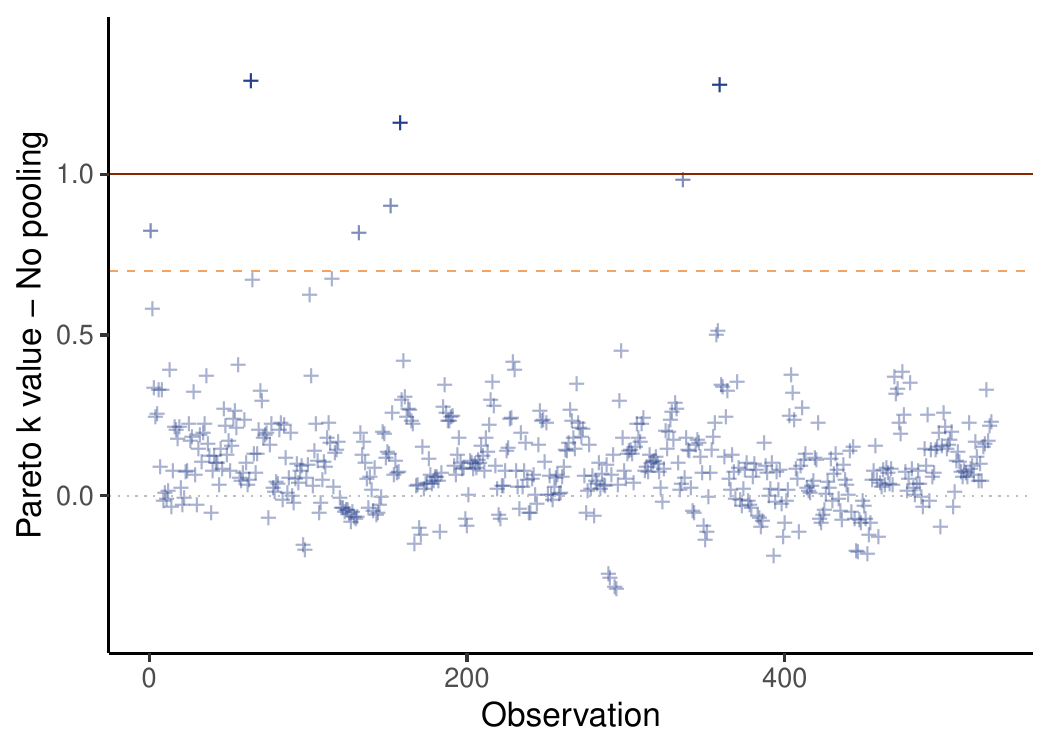}
    \end{subfigure}
        \begin{subfigure}[tb]{.49\linewidth}
      \centering
      \includegraphics[width=\linewidth]{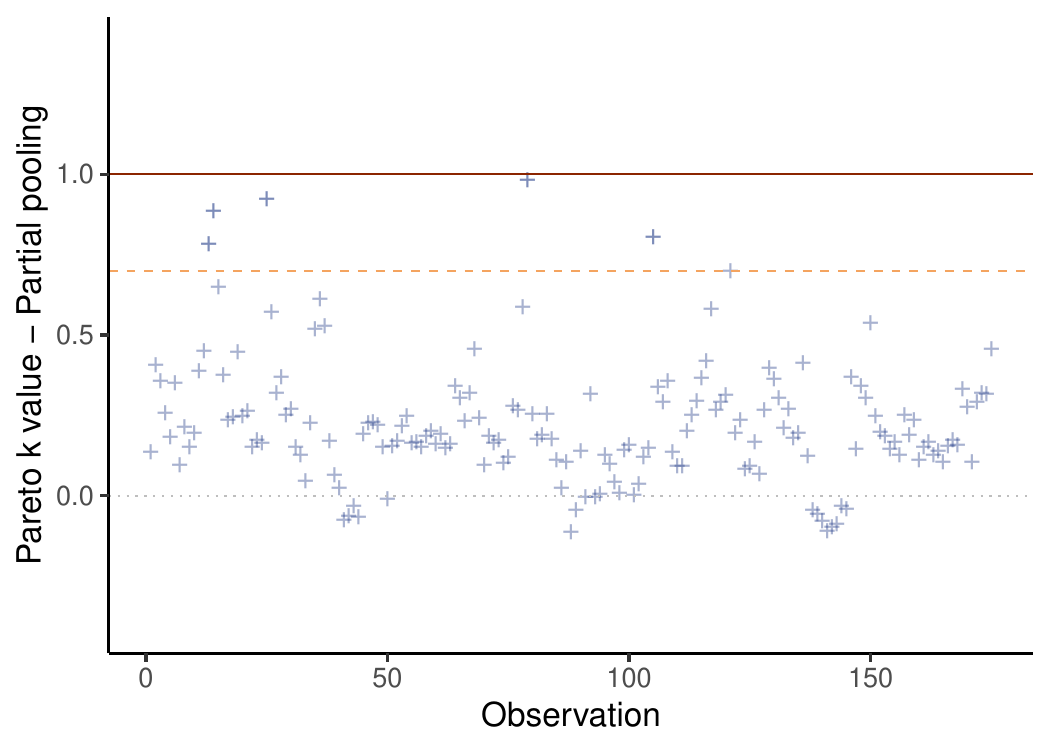}
    \end{subfigure}
    \hfill
    \begin{subfigure}[tb]{.49\linewidth}
      \centering
      \includegraphics[width=\linewidth]{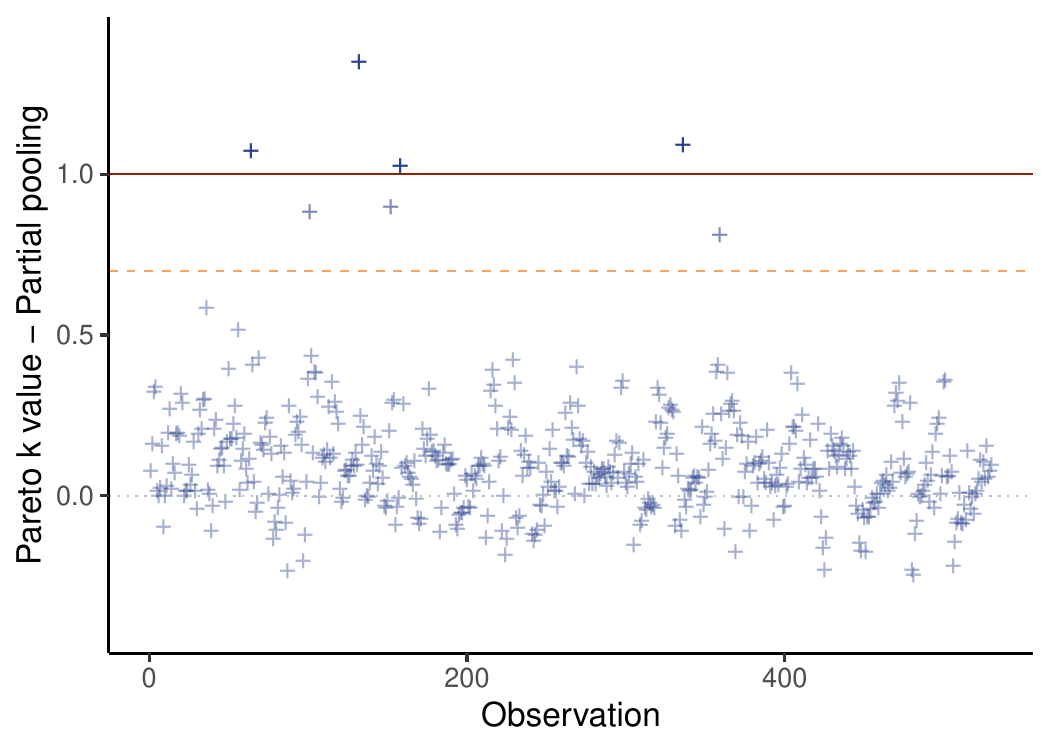}
    \end{subfigure}
        \begin{subfigure}[tb]{.49\linewidth}
      \centering
      \includegraphics[width=\linewidth]{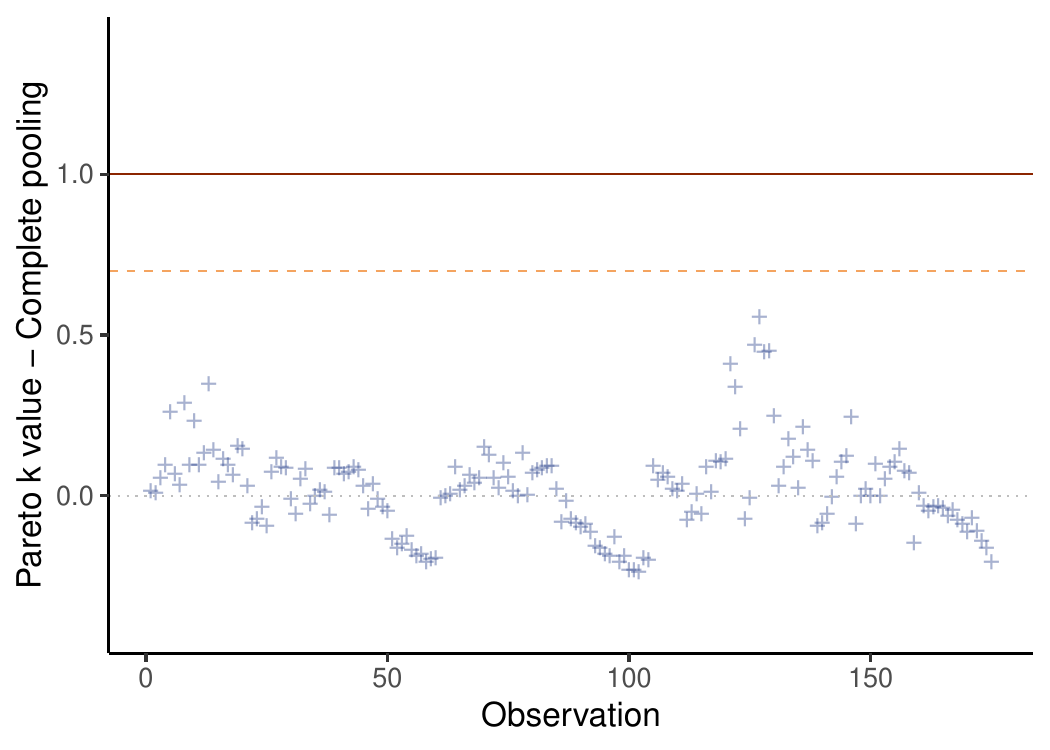}
    \end{subfigure}
    \hfill
    \begin{subfigure}[tb]{.49\linewidth}
      \centering
      \includegraphics[width=\linewidth]{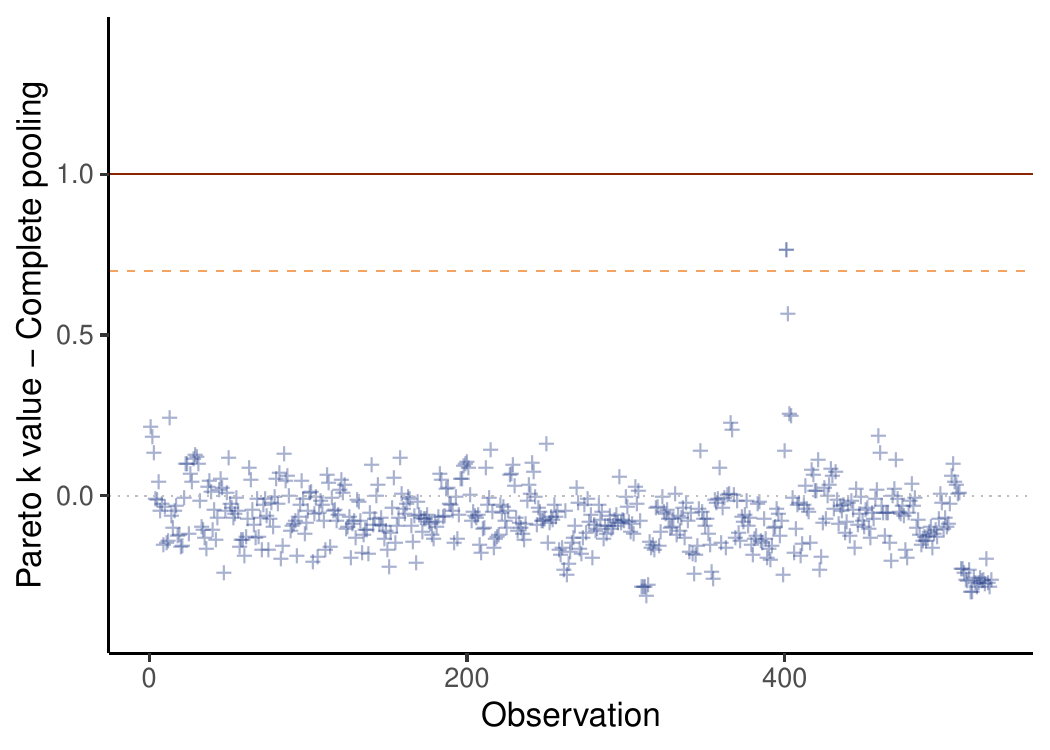}
    \end{subfigure}
    \caption{\textbf{Prostate cancer biomarker example cont.}  
    Diagnostic Pareto $k$ estimates for observations corresponding to (left) the first treatment cycle, and (right) all but the last cycle. Estimates above $k=0.7$ (orange line) are considered high, and estimates above $k=1.0$ (red line) are considered very high.}\label{fig:biomarker-supp-3}
  \end{figure}

\noindent Fig.~\ref{fig:biomarker-supp-3} shows the Pareto $k$ diagnostic plots obtained when computing the $\widehat{\mathrm{elpd}}_{\mathrm{loo}}$ value for the biomarker dataset. In some cases, the observations correspond to high ($k>0.7$) or very high ($k>1.0$) estimates. 
Here, this is often recorded as the first predicted value for each patient and, as such, is likely due to the assumed initial condition for the non-observed states $S, D$. 
Other times, high $k$ estimates correspond to the sudden rise and fall of PSA concentration experienced during, and just after, treatment administration.

\vspace{1.cm}

\newpage

\let\oldthebibliography\thebibliography
\let\endoldthebibliography\endthebibliography
\renewenvironment{thebibliography}[1]{
  \begin{oldthebibliography}{#1}
    \setlength{\itemsep}{0em}
    \setlength{\parskip}{0em}
}
{
  \end{oldthebibliography}
}

\begin{spacing}{0.9}
\renewbibmacro{in:}{,}
\section*{References}
\AtNextBibliography{\small}
\printbibliography[heading=none]
\end{spacing}